\documentclass[reqno,12pt,a4paper]{amsart}

\usepackage{mathrsfs}
\usepackage{amssymb}
\usepackage{amsfonts}
\usepackage{latexsym}
\usepackage{amsthm}

\usepackage{graphicx}
\def\lb{\label}

\newcommand{\er}[1]{\textrm{(\ref{#1})}}

\begin{document}


\renewcommand{\theequation}{\arabic{section}.\arabic{equation}}
\theoremstyle{plain}
\newtheorem{theorem}{\bf Theorem}[section]
\newtheorem{lemma}[theorem]{\bf Lemma}
\newtheorem{corollary}[theorem]{\bf Corollary}
\newtheorem{proposition}[theorem]{\bf Proposition}
\newtheorem{definition}[theorem]{\bf Definition}
\newtheorem{condition}[theorem]{\bf Condition}
\newtheorem{remark}[theorem]{\bf Remark}

\def\a{\alpha}  \def\cA{{\mathcal A}}     \def\bA{{\bf A}}  \def\mA{{\mathscr A}}
\def\b{\beta}   \def\cB{{\mathcal B}}     \def\bB{{\bf B}}  \def\mB{{\mathscr B}}
\def\g{\gamma}  \def\cC{{\mathcal C}}     \def\bC{{\bf C}}  \def\mC{{\mathscr C}}
\def\G{\Gamma}  \def\cD{{\mathcal D}}     \def\bD{{\bf D}}  \def\mD{{\mathscr D}}
\def\d{\delta}  \def\cE{{\mathcal E}}     \def\bE{{\bf E}}  \def\mE{{\mathscr E}}
\def\D{\Delta}  \def\cF{{\mathcal F}}     \def\bF{{\bf F}}  \def\mF{{\mathscr F}}
\def\c{\chi}    \def\cG{{\mathcal G}}     \def\bG{{\bf G}}  \def\mG{{\mathscr G}}
\def\z{\zeta}   \def\cH{{\mathcal H}}     \def\bH{{\bf H}}  \def\mH{{\mathscr H}}
\def\e{\eta}    \def\cI{{\mathcal I}}     \def\bI{{\bf I}}  \def\mI{{\mathscr I}}
\def\p{\psi}    \def\cJ{{\mathcal J}}     \def\bJ{{\bf J}}  \def\mJ{{\mathscr J}}
\def\vT{\Theta} \def\cK{{\mathcal K}}     \def\bK{{\bf K}}  \def\mK{{\mathscr K}}
\def\k{\kappa}  \def\cL{{\mathcal L}}     \def\bL{{\bf L}}  \def\mL{{\mathscr L}}
\def\l{\lambda} \def\cM{{\mathcal M}}     \def\bM{{\bf M}}  \def\mM{{\mathscr M}}
\def\L{\Lambda} \def\cN{{\mathcal N}}     \def\bN{{\bf N}}  \def\mN{{\mathscr N}}
\def\m{\mu}     \def\cO{{\mathcal O}}     \def\bO{{\bf O}}  \def\mO{{\mathscr O}}
\def\n{\nu}     \def\cP{{\mathcal P}}     \def\bP{{\bf P}}  \def\mP{{\mathscr P}}
\def\r{\rho}    \def\cQ{{\mathcal Q}}     \def\bQ{{\bf Q}}  \def\mQ{{\mathscr Q}}
\def\s{\sigma}  \def\cR{{\mathcal R}}     \def\bR{{\bf R}}  \def\mR{{\mathscr R}}
\def\S{\Sigma}  \def\cS{{\mathcal S}}     \def\bS{{\bf S}}  \def\mS{{\mathscr S}}
\def\t{\tau}    \def\cT{{\mathcal T}}     \def\bT{{\bf T}}  \def\mT{{\mathscr T}}
\def\f{\phi}    \def\cU{{\mathcal U}}     \def\bU{{\bf U}}  \def\mU{{\mathscr U}}
\def\F{\Phi}    \def\cV{{\mathcal V}}     \def\bV{{\bf V}}  \def\mV{{\mathscr V}}
\def\P{\Psi}    \def\cW{{\mathcal W}}     \def\bW{{\bf W}}  \def\mW{{\mathscr W}}
\def\o{\omega}  \def\cX{{\mathcal X}}     \def\bX{{\bf X}}  \def\mX{{\mathscr X}}
\def\x{\xi}     \def\cY{{\mathcal Y}}     \def\bY{{\bf Y}}  \def\mY{{\mathscr Y}}
\def\X{\Xi}     \def\cZ{{\mathcal Z}}     \def\bZ{{\bf Z}}  \def\mZ{{\mathscr Z}}
\def\O{\Omega}
\def\th{\theta}

\newcommand{\gA}{\mathfrak{A}}
\newcommand{\gB}{\mathfrak{B}}
\newcommand{\gC}{\mathfrak{C}}
\newcommand{\gD}{\mathfrak{D}}
\newcommand{\gE}{\mathfrak{E}}
\newcommand{\gF}{\mathfrak{F}}
\newcommand{\gG}{\mathfrak{G}}
\newcommand{\gH}{\mathfrak{H}}
\newcommand{\gI}{\mathfrak{I}}
\newcommand{\gJ}{\mathfrak{J}}
\newcommand{\gK}{\mathfrak{K}}
\newcommand{\gL}{\mathfrak{L}}
\newcommand{\gM}{\mathfrak{M}}
\newcommand{\gN}{\mathfrak{N}}
\newcommand{\gO}{\mathfrak{O}}
\newcommand{\gP}{\mathfrak{P}}
\newcommand{\gQ}{\mathfrak{Q}}
\newcommand{\gR}{\mathfrak{R}}
\newcommand{\gS}{\mathfrak{S}}
\newcommand{\gT}{\mathfrak{T}}
\newcommand{\gU}{\mathfrak{U}}
\newcommand{\gV}{\mathfrak{V}}
\newcommand{\gW}{\mathfrak{W}}
\newcommand{\gX}{\mathfrak{X}}
\newcommand{\gY}{\mathfrak{Y}}
\newcommand{\gZ}{\mathfrak{Z}}

\newcommand{\gm}{\mathfrak{m}}
\newcommand{\gn}{\mathfrak{n}}
\newcommand{\gf}{\mathfrak{f}}
\newcommand{\gh}{\mathfrak{h}}
\newcommand{\mg}{\mathfrak{g}}
\newcommand{\gb}{\mathfrak{b}}
\newcommand{\gu}{\mathfrak{u}}
\newcommand{\ga}{\mathfrak{a}}
\newcommand{\gv}{\mathfrak{v}}
\newcommand{\gc}{\mathfrak{c}}
\newcommand{\gj}{\mathfrak{j}}

\def\ve{\varepsilon}   \def\vt{\vartheta}    \def\vp{\varphi}    \def\vk{\varkappa}

\def\Z{{\mathbb Z}}    \def\R{{\mathbb R}}   \def\C{{\mathbb C}}    \def\K{{\mathbb K}}
\def\T{{\mathbb T}}    \def\N{{\mathbb N}}   \def\dD{{\mathbb D}}


\def\la{\leftarrow}              \def\ra{\rightarrow}            \def\Ra{\Rightarrow}
\def\ua{\uparrow}                \def\da{\downarrow}
\def\lra{\leftrightarrow}        \def\Lra{\Leftrightarrow}


\def\lt{\biggl}                  \def\rt{\biggr}
\def\ol{\overline}               \def\wt{\widetilde}
\def\no{\noindent}


\let\ge\geqslant                 \let\le\leqslant
\def\lan{\langle}                \def\ran{\rangle}
\def\/{\over}                    \def\iy{\infty}
\def\sm{\setminus}               \def\es{\emptyset}
\def\ss{\subset}                 \def\ts{\times}
\def\pa{\partial}                \def\os{\oplus}
\def\om{\ominus}                 \def\ev{\equiv}
\def\iint{\int\!\!\!\int}        \def\iintt{\mathop{\int\!\!\int\!\!\dots\!\!\int}\limits}
\def\el2{\ell^{\,2}}             \def\1{1\!\!1}
\def\sh{\sharp}
\def\wh{\widehat}
\def\bs{\backslash}

\def\sh{\mathop{\mathrm{sh}}\nolimits}
\def\Area{\mathop{\mathrm{Area}}\nolimits}
\def\arg{\mathop{\mathrm{arg}}\nolimits}
\def\const{\mathop{\mathrm{const}}\nolimits}
\def\det{\mathop{\mathrm{det}}\nolimits}
\def\diag{\mathop{\mathrm{diag}}\nolimits}
\def\diam{\mathop{\mathrm{diam}}\nolimits}
\def\dim{\mathop{\mathrm{dim}}\nolimits}
\def\dist{\mathop{\mathrm{dist}}\nolimits}
\def\Im{\mathop{\mathrm{Im}}\nolimits}
\def\Iso{\mathop{\mathrm{Iso}}\nolimits}
\def\Ker{\mathop{\mathrm{Ker}}\nolimits}
\def\Lip{\mathop{\mathrm{Lip}}\nolimits}
\def\rank{\mathop{\mathrm{rank}}\limits}
\def\Ran{\mathop{\mathrm{Ran}}\nolimits}
\def\Re{\mathop{\mathrm{Re}}\nolimits}
\def\Res{\mathop{\mathrm{Res}}\nolimits}
\def\res{\mathop{\mathrm{res}}\limits}
\def\sign{\mathop{\mathrm{sign}}\nolimits}
\def\span{\mathop{\mathrm{span}}\nolimits}
\def\supp{\mathop{\mathrm{supp}}\nolimits}
\def\Tr{\mathop{\mathrm{Tr}}\nolimits}
\def\BBox{\hspace{1mm}\vrule height6pt width5.5pt depth0pt \hspace{6pt}}
\def\as{\text{as}}
\def\all{\text{all}}
\def\where{\text{where}}
\def\Dom{\mathop{\mathrm{Dom}}\nolimits}


\newcommand\nh[2]{\widehat{#1}\vphantom{#1}^{(#2)}}
\def\dia{\diamond}

\def\Oplus{\bigoplus\nolimits}



\def\qqq{\qquad}
\def\qq{\quad}
\let\ge\geqslant
\let\le\leqslant
\let\geq\geqslant
\let\leq\leqslant
\newcommand{\ca}{\begin{cases}}
\newcommand{\ac}{\end{cases}}
\newcommand{\ma}{\begin{pmatrix}}
\newcommand{\am}{\end{pmatrix}}
\renewcommand{\[}{\begin{equation}}
\renewcommand{\]}{\end{equation}}
\def\eq{\begin{equation}}
\def\qe{\end{equation}}
\def\[{\begin{equation}}
\def\bu{\bullet}
\def\ced{\centerdot}
\def\tes{\textstyle}

\newcommand{\fr}{\frac}
\newcommand{\tf}{\tfrac}

\title[Inverse problem for the divisor of the good Boussinesq equation]
{Inverse problem for the divisor of the good Boussinesq equation}

\date{\today}
\author[Andrey Badanin]{Andrey Badanin}
\address{Saint-Petersburg
State University, Universitetskaya nab. 7/9, St. Petersburg,
199034 Russia, an.badanin@gmail.com, a.badanin@spbu.ru}
\author[Evgeny Korotyaev]{Evgeny L. Korotyaev}
\address{Academy for Advanced interdisciplinary Studies
Northeast Normal University,
Changchun, 130024 Jilin, China,
korotyaev@gmail.com}


\maketitle

\begin{abstract}
A third-order operator with periodic coefficients is an L-operator
in the Lax pair for the Boussinesq equation on a circle.
The projection of the divisor of the Floquet solution poles
for this operator coincides with the spectrum of the three-point
Dirichlet problem. The sign of the norming constant of
the three-point problem determines the sheet of the Riemann surface
on which the pole lies. We solve the inverse problem for
a third-order operator with three-point Dirichlet conditions
when the spectrum and norming constant are known. We construct
a mapping from the set of coefficients to the set of spectral data
and prove that this mapping
is an analytic bijection in the neighborhood of zero.
\end{abstract}

\begin{quotation}
\begin{center}
{\bf Table of Contents}
\end{center}

\vskip 6pt

{\footnotesize

1. Introduction and main results \hfill \pageref{Sec1}\ \ \ \ \

2. Monodromy matrices and their properties  \hfill \pageref{Sec2}\ \ \ \ \

3. The fundamental matrix at the three-point eigenvalues  \hfill \pageref{Sec3}\ \ \ \ \

4. Eigenfunctions  \hfill \pageref{Sec5}\ \ \ \ \

5. Gradient of the characteristic function $\Delta$ \hfill \pageref{Sec4}\ \ \ \ \

6.  Gradients of the three-point eigenvalues \hfill \pageref{Sec6}\ \ \ \ \

7. Gradients of the norming constants  \hfill \pageref{Sec7}\ \ \ \ \

8. Asymptotics of the gradients of the norming constants
\hfill \pageref{Sec8}\ \ \ \ \

 }
\end{quotation}

\section{\lb{Sec1} Introduction and main results}
\setcounter{equation}{0}

\subsection{Introduction}
 We consider an inverse problem for a non-self-adjoint
operator $\cL=\cL(p,q)$ acting on $L^2(0,2)$ and given by
\[
\lb{Hdpq}
\cL y=(y''+py)'+py'+qy,\qqq y(0)=y(1)=y(2)=0,
\]
where the 1-periodic coefficients $p,q$ belong to the real Hilbert space
\[
\lb{defHs}
\cH=\{f\in L_\R^2(\T,\R):\int_0^1f(x)dx=0\},\qq \T=\R/\Z,
\]
equipped with the norm $\|f\|^2=\int_0^1|f(x)|^2dx$.
The spectrum of $\cL $ is pure discrete.

Many works are dedicated to multi-point boundary value problems.
For more information, see the review in our paper \cite{BK21}.
In the theory of finite-zone integration, multi-point problems
arose in connection with the use of the Baker-Akhiezer function.
For scalar operators
of order $N\ge 3$ and for $N\ts N$ matrix operators of first order,
the poles of the Baker-Akhiezer
function are projected onto the eigenvalues of the $N$-point problem
(see references \cite{McK81}, \cite{D85}, \cite{P03}).
For the Schr\"odinger operator,
this problem is the standard two-point Dirichlet problem.
The projections of the poles of the Baker-Akhiezer function
for our third-order operator
are the eigenvalues of the operator $\cL$.

The spectrum of our operator $\cL$ is the set of projections
of the Baker-Akhiezer function's
poles for the {\it good Boussinesq equation}
$$
\tes
p_{tt}=-{1\/3}(p_{xxxx}+4(p^2)_{xx}), \qqq p_t=q_x,
$$
on the circle.
To describe these poles, we consider the differential equation
\[
\lb{1b}
(y''+py)'+py'+q y=\l y,\qqq \l\in\C.
\]
Let $ y_1,  y_2,  y_3$ be the fundamental solutions
to Eq.~\er{1b}
under the conditions
$ y_j^{[k-1]}|_{x=0}=\d_{jk},j,k=1,2,3$, where
$y^{[0]}=y$, $y^{[1]}=y'$, $y^{[2]}=y''+ py$.
Each of the function $ y_j(x,\cdot),j=1,2,3,x\in\R$, is entire
and real on $\R$. Define the {\it monodromy matrix} $M$ and its characteristic determinant $D$ by
\[
\lb{defmm}
M(\l)=\big( y_j^{[k-1]}(1,\l)\big)_{j,k=1}^3,
\]
\[
\lb{1c}
D(\t,\l)=\det(M(\l)-\t \1_{3}),\qq (\t,\l)\in\C^2.
\]
The matrix-valued function $M$ and the function $D$ are entire.
An eigenvalue  of $M$ is called a {\it multiplier}, it is a
zero of the polynomial $D(\cdot,\l)$.
The matrix $M$ has exactly $3$ (counting with
multiplicities) eigenvalues $\t_j,j=1,2,3$, which  satisfy
$
\t_1\t_2\t_3=1.
$
In particular, each $\t_j\ne 0$ for all $\l\in\C$.
The functions $\t_1,\t_2$, and $\t_3$ constitute 3 branches of an analytic
function on a 3-sheeted Riemann surface $\cR$ having only algebraic singularities.
This surface is an invariant of the Boussinesq flow.
For each multiplier $\t_j,j=1,2,3$, there exists the Floquet solution
$\gf_j(x,\l),(x,\l)\in\R\ts\C$,
to Eq.~\er{1b} such that
$$
\gf_j(x+1,\l)=\t_j(\l)\gf_j(x,\l),\qq \gf_j(0,\l)=1.
$$
For each $x\in\R$ the functions $\gf_j(x,\cdot),j=1,2,3$,
constitute 3 branches of {\it the Baker-Akhiezer function} $\gf(x,\cdot)$,
meromorphic on the surface $\cR$.
If $\l$ is not a branch point of $\cR$, and the monodromy matrix $M(\l)$
has the eigenvector of the form $(0,a,b)^\top,a,b\in\C$, corresponding
to the multiplier $\t_j$ for some $j=1,2,3$, then $\l$ is a pole of
the function $\gf_j(x,\cdot)$ for all $x\in\R$. The set of poles of the function
$\gf(x,\cdot)$ on $\cR$ form the {\it pole divisor}
of the function $\gf$.
McKean \cite{McK81} obtained  following results:

\no  {\it
Let the multiplier surface $\cR$ be fixed.
The spectral class of pairs of sufficiently small
$(p, q)\in C^\iy\ts C^\iy$
producing this surface, is considered.
The map from the pair $(p, q)$
in this class to the pole divisor of the
Baker-Akhiezer function is one-to-one.
}

\medskip

Dickson, Gesztesy, and Unterkofler \cite{DGU99x}
provide an explicit expression for the solution to the Boussinesq equation
in terms of the theta function of the Riemann surface $\cR$
in the ``finite-gap case.'' This formula is similar
to the formula that Its and Matveev provided for the Korteweg-de Vries
equation (see \cite{IM75}).
In this expression, the divisor fully determines the dynamics of the solution.
One of our goals in this and the next study is to lay the foundation
for extending the Dickson-Gesztesy-Unterkofler formula
to the ``infinite-gap case.''

As mentioned earlier,
the set of projections of the poles onto the complex plane
is the same as the spectrum of the operator $\cL$.
The surface sheet on which a pole is located is determined
by the sign of its corresponding norming constant,
at least for  small coefficients $p$ and $q$
(see \cite{BK24x}). In this paper, we define
a mapping from the set of pairs of coefficients $(p,q)$
in space $\cH\ts\cH$
to the set of pairs of eigenvalues and their corresponding norming constants.
We prove that this mapping
is an analytic bijection in the neighborhood of the point $p=q=0$.

McKean \cite{McK81} discusses the operator $\cL$
in the case of infinitely smooth, small coefficients from $C^\iy(\T)$.
In contrast to McKean, here we consider the problem in certain classes
of smoothness of coefficients. This problem is more difficult
than the one McKean considered. However, like McKean,
we also consider small coefficients.
This case is simpler because the eigenvalues are simple.

Ultimately, our results show that a solution to the Boussinesq
equation with sufficiently small initial data from a fixed smoothness
class exists globally within that class. However, if the coefficients
of the operator are not small, the eigenvalues may be multiples.
This complicates the analysis and results in solutions
to the Boussinesq equation exhibiting blow-up,
as discovered by Kalantarov and Ladyzhenskaja \cite{KL77}.
The corresponding inverse problem remains unsolved.

In this paper, we use our results from \cite{BK25} a lot. In that paper,
we studied the direct spectral problem when the coefficients are small.
We found uniform energy asymptotics of the fundamental solutions,
multipliers, eigenvalues, and norming constants.

The main results of the present article are outlined in \cite{BK24xx}.
Our proofs are based on the methods of P\"oschel and Trubowitz \cite{PT87}
and Korotyaev \cite{K99}.
The mapping in our study of {\it the inverse three-point problem}
can be expressed as a composition of two mappings.
The first is the McKean transformation, which converts
the three-point problem of a third-order equation into
a Dirichlet problem for a Schrodinger equation
with an energy-dependent potential. The McKean transformation
was studied in the article \cite{BK24x}, and
the Schrodinger equation with an energy-dependent potential
was considered in the article \cite{BK21x}. The second mapping is
from the set of potentials of the Schrodinger operator to
the set of pairs of eigenvalues and their corresponding norming constants
for the Dirichlet problem. Such a mapping was constructed and studied
in the book by P\"oschel and Trubowitz \cite{PT87}.

The results obtained here are important for
solving {\it the inverse periodic problem}.
An outline of the solution to this problem is presented in \cite{BK24xxx}.
In that paper, the mapping from the set of coefficients to the set
of spectral data for a periodic problem is a composition of two mappings:
the McKean transformation and the mapping from the set of Hill operator potentials
to the set of spectral data.
Korotyaev defined the corresponding mapping in \cite{K98},
\cite{K99}.

Many works have been devoted to inverse spectral problems for third-order operators.
Articles by Amour \cite{A99}, \cite{A01},
Bondarenko \cite{Bo23}, Liu and Yan \cite{LY23}, and Zolotarev \cite{Zo24}
have studied inverse problems for third-order operators
on a finite interval under various boundary conditions.
Deift, Tomei, and Trubowitz studied the inverse scattering
problem for third-order operators in their article \cite{DTT82}.

Describe our main results.
Introduce the vector-valued functions
$\gu,\gu_*,\gu^-,\gu_*^-\in \gH=\cH\os\cH$ by
\[
\lb{epsstar}
\gu=(p,q),\qq \gu_*=(p,-q),\qq
\gu^-(x)=\gu(1-x),\qq
\gu_*^-(x)=\gu_*(1-x),\ \ x\in\R.
\]
Introduce the Hilbert spaces
\[
\lb{defspgH}
\cH_1=\{f:f,f'\in\cH\},\qq
\gH=\cH\os\cH,\qq
\gH_1=\cH_1\os\cH,
\]
equipped with the norms
$$
\|\gu\|^2=\|p\|^2+\|q\|^2,\qq
\|\gu\|_1^2=\|p'\|^2+\|q\|^2,
$$
where $\cH$ is given by \er{defHs}.
Let $\gH_\C$ be the complexification of the real space $\gH$ and let
$\gH_{1,\C}$ be the complexification of $\gH_1$.
 Define balls $\cB(r),r>0$, in the space $\gH_1$  by
$$
\cB(r)=\{\gu\in\gH_1:\|\gu\|_1<r\}, \qqq r>0,
$$
and the corresponding balls  $\cB_\C(r)$ in $\gH_{1,\C}$.

{\it We will construct a mapping from the ball $(p,q)\in \cB(\ve)$ into spectral data, here and throughout the text,
$\ve$ is a well-defined positive number that is sufficiently small,
a certain constant.}

\subsection{Eigenvalues}
The spectrum $\s(\cL )$ of the operator $\cL $ is pure discrete
and satisfies
\[
\lb{spec}
\s(\cL )=\{\l\in\C:\Delta(\l)=0\},
\]
where $\Delta$ is an entire function given by
\[
\lb{defsi}
\Delta(\l)=\det\ma y_2(1,\l)& y_3(1,\l)\\
 y_2(2,\l)& y_3(2,\l)\am.
\]
The operators $\cL(\gu)$ and $-\cL(\gu_*^-)$
are unitarily equivalent,
then
$
\s\big(\cL (\gu)\big)=-\s\big(\cL (\gu_*^-)\big).
$

In the unperturbed case $\gu=0$,
all eigenvalues are simple and real. They have the form
\[
\lb{unpev}
\tes
\m_{n}^o=(2\nu n)^3,\qq
n\in\Z_0:=\Z\sm\{0\},\qq \nu={\pi\/\sqrt3}.
\]
Introduce domains $\cD_n$ by
\[
\lb{DomcD}
\cD_{n}=\{\l\in\C:|z-2\nu n|
<1\},\qq\cD_{-n}=\{\l\in\C:-\l\in\cD_n\},\qq n\ge 0,
\]
here and below
\[
\lb{S} \tes
 z=\l^{1\/3},\qq\arg z\in\big(-{\pi\/3},{\pi\/3}\big],
\qq  \arg\l\in(-\pi,\pi].
\]
We proved in \cite{BK25} that
if $\gu\in\cB(\ve)$, then there is exactly one simple real zero $\m_n$
 of the function
$\Delta$ in each domain $\cD_n,n\in\Z_0$, and there are no other zeros in $\C$.
So, in this case all spectrum is real.
Moreover, the eigenvalues $\mu_n$ satisfy
\[
\lb{symev}
\m_{-n}(\gu)=-\m_n(\gu_*^-)\qq\forall\ \ n\in\Z_0.
\]

In order to formulate our first theorem about the eigenvalues
 we introduce the Fourier transform
\[
\lb{defwhtfcns}
\mF_{cn}f=\wh f_{cn}=\int_0^1f(x)\cos 2\pi nxdx,\ \
\mF_{sn}f=\wh f_{sn}=\int_0^1f(x)\sin 2\pi nxdx,\ \ n\in\N.
\]

\begin{theorem}
\lb{Th3pram}
Each eigenvalue $\mu_n, n\in\Z_0$, is analytic in the ball
$\cB_\C(\ve)$. Moreover,
if $\gu\in\cB(\ve)$ and  $n\in\Z_0$, then
\[
\lb{as3pev}
\Big|h_{cn}(\gu)+{\wh p_{sn}'\/\sqrt3}+\wh q_{cn}
-{\wh p_{cn}'\/3}+{\wh q_{sn}\/\sqrt3}\Big|
\le {C\|\gu\|_1^2\/|n|},
\]
where
\[
\lb{defhcn}
h_{cn}=\m_n-\m_n^o.
\]
The gradient $ {\pa\m_n\/\pa\gu(t)} =\big({\pa\m_n\/\pa
p(t)},{\pa\m_n\/\pa q(t)}\big) $ satisfies
\[
\lb{aspamun}
\begin{aligned}
\max\Big\{\Big|{\pa\m_n(\gu)\/\pa p(t)}
-{2\pi n\/3}b_{2n}(t)\Big|,\
\Big|{\pa\m_n(\gu)\/\pa q(t)}
+{b_{2n}(t)\/\sqrt3}\Big|\Big\}
\le {C\|\gu\|_1\/|n|},\qqq t\in[0,1],
\end{aligned}
\]
for some $C>0$, where $b_{2n}(t)=\sqrt3\cos(2\pi nt)+\sin(2\pi nt)$.
\end{theorem}

\no {\bf Remark.}
The estimate \er{as3pev} was proved in our paper \cite{BK25}.

\subsection{Norming constant}
Introduce the domain
\[
\lb{defmD3}
\mD=\C\sm\cup_{n\in\{0\}\cup\N}\ol{\cD_{-n}}.
\]
We proved in \cite[Lm~4.4]{BK25} that if
$(\l,\gu)\in\mD\ts\cB(\ve)$,
then the multiplier $\t_3(\l)$ is simple and satisfies
\[
\lb{lom}
|\t_3(\l)-e^{z}|\le {C\|\gu\|_1^2\/|z|^4}|e^{z}|,
\]
for some $C>0$.
The function $\t_3$ is analytic on the domain $\mD$,
real for real $\l$, and $\t_3(\l)>0$ for all
$\l>1$. Moreover, each $\t_3(\l,\cdot),\l\in\mD$, is
analytic on the ball $\cB_\C(\ve)$.

Since we are solving the inverse problem for a non-self-adjoint operator,
we have to study both the operator $\cL$
 and the transposed (in McKean's terminology) operator
\[
\lb{1btr}
\wt\cL \wt y=-(\tilde y''+p\tilde y)'-p\tilde y'+q \tilde y,\qqq
\tilde y(0)=\tilde y(1)=\tilde y(2)=0.
\]
If  $\gu\in\cB(\ve)$, then in each
domain $\cD_n,n\in\Z_0$, there is exactly one real eigenvalue $\wt\m_n$ of
the operator $\wt\cL$, moreover, $\wt\m_n=-\m_{-n}(\gu_*)$, where
$\gu_*$ is given by \er{epsstar}. There are no other eigenvalues in $\C$.
Let $\wt y_n(x)$ be the corresponding eigenfunction such that $\wt y_n'(0)=1$.
We define the norming constants $h_{s,\pm n}, n\in \N$, by
\[
\lb{defnf}
\begin{aligned}
& h_{s,n} =8(\pi n)^2\log |\wt y_n'(1)\t_3^{-{1\/2}}(\wt\m_n)|, \qq
\t_3^{1\/2}(\wt\m_n)>0,\qq
\\
& h_{s,-n}(\gu)=-h_{s,n}(\gu_*^-),
\end{aligned}\qq n\in\N.
\]
where $\gu_*^-$ is given by \er{epsstar}.

In \cite{BK24x} we consider the transformation, constructed by McKean
\cite{McK81}. This transformation reduces the 3-point Dirichlet problem \er{1btr}
for our transpose operator $\wt\cL$ to the standard 2-point Dirichlet problem for
the Schr\"odinger operator with an energy-dependent potential
$V(x,\l)$. Moreover, McKean's transformation maps:

1) The set of the
eigenvalues $\wt\m_n$ of the 3-point Dirichlet problem for the transpose
operator onto the set of the Dirichlet eigenvalues for
the Schr\"odinger operator,

2) The set of the norming constants $h_{s,n}$, given by \er{defnf},
onto the set of the norming constants of the Dirichlet problem
for the Schr\"odinger operator, defined in \cite[Ch~3]{PT87}.

3) The set of branch points of the surface $\cR$ onto
the set of the two-periodic eigenvalues for the
Schr\"odinger operator.

\begin{theorem}
\lb{Thnf} Each function $h_{sn},n\in\Z_0$, is analytic on the ball
$\cB_\C(\ve)$. Moreover, if $\gu\in\cB(\ve)$, then
\[
\lb{asncr}
\Big|h_{sn}(\gu)
-\Big({\wh p_{sn}'\/\sqrt3}+\wh q_{cn}
-\wh p_{cn}'+\sqrt3\wh q_{sn}\Big)\Big|\le{ C\|\gu\|_1^2\/|n|},
\]
\[
\lb{asgrhcn}
\begin{aligned}
\max\Big\{
\Big|{\pa h_{sn}(\gu)\/\pa p(t)}
-{2\pi n\/\sqrt3}a_{-2n}(t)\Big|,\
\Big|{\pa h_{sn}(\gu)\/\pa q(t)}
+a_{-2n}(t)\Big|\Big\}
\le{C\|\gu\|_1\/|n|},
\end{aligned}
\]
for all $n\in\Z_0$ and for some $C>0$, where
$a_{2n}(t)=\sqrt3\sin(2\pi nt)-\cos(2\pi nt)$.
\end{theorem}

\no {\bf Remark.}
The estimates \er{asncr} are obtained in \cite{BK25}.

\subsection{Inverse results}
Introduce the mapping $h:\gu\to h(\gu)$ on $\cB(\ve)$ by
\[
\lb{defh}
h(\gu)=(h_n(\gu))_{n\in\Z_0},
\]
where the vector $h_n:\gH_1\to\R^2$ has the form
\[
\lb{defhn} h_n=\big(h_{cn},h_{sn}\big),
\]
the first component $h_{cn}$ is given by \er{defhcn} and
the second component $h_{sn}$ is given by \er{defnf}.
The asymptotics \er{as3pev} and \er{asncr} show that
$h(\gu)\in\ell^2\os\ell^2$, where
$
\ell^2=\{(a_n)_{n\in\Z_0}:\sum_{n\in\Z_0}|a_n|^2<\iy\}.
$
Introduce the linear isomorphism  $\cF:\gH_1\to\ell^2\os\ell^2$ by
\[
\lb{defcF}
\cF\gu=(\cF_n\gu)_{n\in\Z_0}.
\]
where the linear transformations $\cF_n:\gH_1\to\C^2,n\in\Z_0$, have
the form
\[
\lb{defFhi}
\cF_{n}\gu=\ma-1&{1\/\sqrt3}\\1&-\sqrt3\am
\ma\mF_{cn}&\mF_{sn}\\-\mF_{sn}&\mF_{cn}\am\ma 0&1\\{1\/\sqrt3}&0\am
\ma p'\\q\am.
\]
We prove the following result.

\begin{theorem}
\lb{ThNablagsn}

The mapping $h:\cB_\C(\ve)\to h(\cB_\C(\ve))$ is a real analytic
bijection between  $\cB_\C(\ve)$ and $h(\cB_\C(\ve))$ and satisfies
\[
\lb{estg-cF}
\|h(\gu)-\cF\gu\|\le C\|\gu\|_1^2,
\]
for all $\gu\in\cB(\ve)$ and for some $C>0$.

\end{theorem}

\no {\bf Remark.}
1) Asymptotics of gradients of the mappings
$h_n$ are given by \er{aspamun} and \er{asgrhcn}.

\no 2) We believe that similar results hold not only
near zero coefficients, but also near any coefficients
for which all the eigenvalues are simple. However,
the situation involving coefficients corresponding
to multiple eigenvalues
is rather complicated and has not yet been studied.

\no 3) If $p$ and $q$ are even, i.e., $u_-=u$, then
$\mu_{-n}(u)=-\mu_{n}(u_*)=\wt\mu_{-n}(u)$.
So in this case the spectra of the operators $\cL$ and $\wt\cL$
are the same.

\medskip

The structure of the paper is as follows: Section~\ref{Sec2}
discusses the fundamental matrix of Eq.~\er{1b}.
We derive properties of the fundamental matrix and prove estimates for it.
Section~\ref{Sec3} continues the discussion
of fundamental solutions and obtains their estimates at the eigenvalues,
which are important for further considerations.
Section~\ref{Sec5} derives properties and estimates for the eigenfunctions.
 Section~\ref{Sec4} considers
the characteristic function of the eigenvalues and calculates its gradient
in Lemma~\ref{Lmgragchf}.
Section~\ref{Sec6} calculates the gradients of the eigenvalues and proves Theorem~\ref{Th3pram}.
The gradients are expressed in terms of the eigenfunctions.
Section~\ref{Sec7} considers the norming constants and calculates their gradients.
Theorems~\ref{Thnf} and \ref{ThNablagsn} are proven in Section~\ref{Sec8}.

\section{ \lb{Sec2} Monodromy matrices and their properties}
\setcounter{equation}{0}

\subsection{ Fundamental matrix}
Let $\gu\in\gH$.
Introduce the $3\ts 3$ {\it fundamental matrix}
$\F(x,\l)$, by
\[
\lb{deM}
\F=(\F_{jk})_{j,k=1}^3=\ma y_1& y_2& y_3\\
 y_1'& y_2'& y_3'\\
 y_1^{[2]}& y_2^{[2]}& y_3^{[2]}\am,\qq \F|_{x=0} =\1_3,\qquad(x,\l)\in\R\ts\C,
\]
where $ y_1,  y_2,  y_3$ are the fundamental solutions to Eq.~\er{1b},
here and below $\1_3$ is the $3\ts 3$ identity matrix, and
\[
\lb{defmL}
y^{[2]}=y''+ py.
\]
The matrix-valued function $\F$, given by \er{deM}, satisfies the equation
\[
\lb{me2}
\F'- \L \F= -V \F,\qq \F|_{x=0}=\1_3,
\]
where
\[
\lb{mtrH}
\L=\ma 0&1&0\\0 &0&1\\\l&0&0\am,\ \
V=\ma 0&0&0\\p&0&0\\q&p&0\am=p J_p+q J_q,\ \
 J_p=\ma 0&0&0\\1&0&0\\0&1&0\am,\ \
 J_q=\ma 0&0&0\\0&0&0\\1&0&0\am.
\]
Each matrix-valued function $\F$
satisfies the Liouville identity
$\det \F=1.$

The matrix $\L$ has the following representation
\[
\lb{mu}
\L=z Z U \O( Z U)^{-1},
\]
where
\[
\lb{4g.Om}
 Z=\diag(1,z,z^2),\qq\O=\diag(\o,\o^2,1),\qqq \o=e^{i{2\pi\/3}},
\]
\[
\lb{defmaZ}
U={1\/\sqrt3}\ma1&1&1\\\o &\o^2 &1 \\
\o^2&\o &1\am,\qq
U^{-1}=U^*={1\/\sqrt3}\ma 1&\o^2&\o\\1&\o&\o^2\\1&1&1\am,\qq
\det U=-i.
\]
For the unperturbed equation $y'''=\l y$, we have
$V=0$. Due to \er{mu}, the corresponding fundamental matrix $\F_0$
has the form
\[
\lb{idX0cU}
\F_0=e^{x\L}= Z U e^{zx\O}( Z U)^{-1}.
\]
The matrix $\L$ has three eigenvalues $\o z,\o^2 z,z$.
Then the eigenvalues of the matrix $\F_0$ have the form
$e^{\o zx},e^{\o^2 zx},e^{zx}$.

Consider the equation
\[
\lb{1btr1}
-(\tilde y''+p\tilde y)'-p\tilde y'+q \tilde y=\l \tilde y
\]
for the transpose operator $\wt\cL$ given by \er{1btr}.
The fundamental matrix $\tilde \F$ of Eq.~\er{1btr1}
has the form
\[
\lb{deMc}
\tilde \F=\ma\tilde y_1&\tilde y_2&\tilde y_3\\
\tilde y_1'&\tilde y_2'&\tilde y_3'\\
\tilde y_1^{[2]}&\tilde y_2^{[2]}&\tilde y_3^{[2]}\am,\qq
\tilde \F|_{x=0} =\1_3,
\]
where $\tilde y_1, \tilde y_2, \tilde y_3$ are the fundamental solutions
to Eq.~\er{1btr1}. Comparing Eqs~\er{1b} and \er{1btr1}, we get
\[
\lb{symXM}
\wt \F(x,\l,\gu)=\F(x,-\l,\gu_*),
\]
where $\gu_*$ is given by \er{epsstar}.
Moreover, the fundamental matrices satisfy
\[
\lb{simMt}
\wt \F^\top J\F=\const=J,\qq \wt \F =J(\F^\top)^{-1}J,\qq\text{where}\ \
J=\ma 0&0&1\\0&-1&0\\1&0&0\am,
\]
see, e.g., \cite{BK25}.

Let $(\l,\gu)\in\C\ts\gH$.
Introduce the fundamental solutions of the equations with shifted coefficients
\[
\lb{defvp3t}
 y_j(x,t,\l,\gu)= y_j(x,\l,\gu(\cdot+t)),\qq
\wt y_j(x,t,\l,\gu)=\wt y_j(x,\l,\gu(\cdot+t)),\qq  j=1,2,3,\qq
(x,t)\in\R^2,
\]
and the fundamental matrix
$\F(x,t,\l,\gu)=\F(x,\l,\gu(\cdot+t))$,
$\wt\F(x,t,\l,\gu)=\wt\F(x,\l,\gu(\cdot+t))$.
The  fundamental matrix
$\F(x,t,\l)=\F(x,t,\l,\gu)$ satisfies
\[
\lb{me2t}
\F'=\big(\L- V(\cdot+t)\big)\F,\qq\F(0,t)=\1_3.
\]
Moreover, $\F(x+t,\l)$ is a solution to Eq.~\er{me2t} satisfying the condition
$\F(x+t,\l)|_{x=0}=\F(t,\l)$, which yields
\[
\lb{shX}
\F(x+t,\l)=\F(x,t,\l)\F(t,\l),\qq (x,t,\l)\in\R^2\ts\C.
\]

\subsection{Derivatives of the fundamental matrix}
We prove the following results about the fundamental matrices.

\begin{lemma}
\lb{T21}
For each $x\in\R$ the matrix-valued function
$\F(x,\l,\gu)$ is an entire function on $\C\ts\gH_\C$.
It is real on $\R\ts\gH$.
Moreover,
for all $(x,\l,\gu)\in[0,2]\ts\C\ts\gH_\C$
and for all $t\in[0,1]$ the following identities hold true:
\[
\lb{derlfm}
\dot \F(x,\l):={\pa \F(x,\l)\/\pa\l}
=\int_0^x \F(x-t,t,\l) J_q\F(t,\l)dt,
\]
\[
\lb{derpsifm}
{\pa \F(x,\l)\/\pa\gu(t)}
=\ca-\Theta(x,t,\l)\1_{[0,x]}(t),x\in[0,1]
\\
-\Theta(x,t,\l)-\Theta(x,t+1,\l)\1_{[0,x-1]}(t),x\in[1,2]\ac,
\]
\[
\lb{grX(1)}
{\pa \F(1,\l)\/\pa\gu(t)}=-\F(1-t,t,\l)( J_p, J_q)\F(t,\l),
\]
\[
\lb{grwtX(1)}
{\pa\wt \F(1,\l)\/\pa\gu(t)}
=-\wt \F(1-t,t,\l)( J_p,- J_q)\wt \F(t,\l),
\]
where $J_p,J_q$ are given by \er{mtrH},
\[
\lb{defcAxt}
\Theta(x,t,\l)=\F(x-t,t,\l)( J_p, J_q)\F(t,\l),
\]
$$
\1_{[0,x]}(t)=\ca 1,t\in[0,x]\\0,t\not\in[0,x]\ac,\qq x>0.
$$

\end{lemma}

\no {\bf Proof.} The standard arguments, see, e.g., \cite[Lm~2.1]{BK14},
show that the matrix-valued function
$\F(x,\cdot,\cdot)$ is entire on $\C\ts\gH_\C$
and real on $\R\ts\gH$.

Let $\gu\in\gH_\C$. Differentiating Eq.~\er{me2} implies
$$
\dot \F'- \L \dot \F+V\dot \F=  J_q \F .
$$
Using $\dot \F|_{x=0}=0$, we obtain
$\dot \F(x)=\int_0^x \F(x)\F^{-1}(t) J_q\F(t)dt$, where $\F(x)=\F(x,\l),...$
The identity \er{shX} implies
$\F(x)\F^{-1}(t)=\F(x-t,t)$. This yields \er{derlfm}.

We prove  \er{derpsifm}. Let $(\l,\gu)\in\C\ts\gH_\C$ be fixed.
Eq.~\er{me2} gives
$$
\F'(x,\gu+\d\gu)- \L \F(x,\gu+\d\gu)+V(x,\gu) \F(x,\gu+\d\gu)= -\d V(x) \F(x,\gu+\d\gu),
$$
where $\d\gu=(\d p,\d q)\in\gH_\C$,
$\d V(x)=V(x,\gu+\d\gu)-V(x,\gu)$, and we write $\F(x,\gu)=\F(x,\l,\gu),...$
The identity \er{mtrH} implies
\[
\lb{prirV}
\d V(x)= J_p\d p(x)+ J_q\d q(x).
\]
This yields
$$
\begin{aligned}
\F(x,\gu+\d\gu)
=\F(x,\gu)-\F(x,\gu)\int_0^x\F^{-1}(t,\gu)\d V(t)\F(t,\gu+\d\gu)dt
\\
=\F(x,\gu)-\F(x,\gu)\int_0^x\F^{-1}(t,\gu)\d V(t)\F(t,\gu)dt+
O(\|\d\gu\|^2),
\end{aligned}
$$
as $\|\d\gu\|\to 0$. The identities \er{shX} and \er{prirV}
give
$$
d_\gu \F(x,\gu)(\d\gu)=-\int_0^x\F(x-t,t,\gu)\big( J_p\d p(t)+ J_q\d q(t)\big)\F(t,\gu)dt
=-\int_0^x\big(\Theta(x,t,\gu),\d\gu(t)\big)dt,
$$
where $\Theta$ is given by \er{defcAxt}. Then
$$
d_\gu \F(x,\gu)(\d\gu)=-\ca
\int_0^x\big(\Theta(x,t,\gu),\d\gu(t)\big)dt,x\in[0,1]\\
\int_0^1\big(\Theta(x,t,\gu),\d\gu(t)\big)dt
+\int_0^{x-1}\big(\Theta(x,t+1,\gu),\d\gu(t)\big)dt,x\in[1,2]
\ac,
$$
which yields \er{derpsifm} and  \er{grX(1)}.
Eq.~\er{symXM} implies
$$
{\pa\wt \F(1,\l,\gu)\/\pa\gu}={\pa \F(1,-\l,\gu_*)\/\pa\gu}
=\Big({\pa \F(1,-\l,p,q)\/\pa p},{\pa \F(1,-\l,p,-q)\/\pa q}\Big).
$$
Then the identity \er{grX(1)} gives \er{grwtX(1)}.~\BBox

\subsection{Asymptotics of the fundamental matrix}
We prove the following simple estimates.

\begin{lemma}
Let $l,m=1,2,3,\l\in\C_+$. Then the following estimates hold true:
\[
\lb{estexpom}
|e^{(\o^m-1)zx}|\le1,
\qq
|e^{(\o-\o^m)zx}|\le 1,\qq x\ge 0,
\]
\[
\lb{estexpomx}
|e^{z(1-x)\o^l+zx\o^m-z}|\le 1,
\qq
|e^{z(x-1)\o^l-zx\o^m+\o z}|\le 1,\qq x\in [0,1].
\]

\end{lemma}

\no {\bf Proof.}
Let $\l\in\C_+$.
Then we have $\Re\o z\le\Re\o^2 z\le\Re z.$
This yields the estimates \er{estexpom}.
Moreover,
\[
\lb{reprom2}
(1-x)\o^l+x\o^m-1=\o^l-1+x(\o^m-\o^l)
=(1-x)(\o^l-\o^m)+\o^m-1.
\]
The second representation in \er{reprom2} gives
the first estimate in \er{estexpomx}
for $1\le m\le l\le 3$, and the third representation in \er{reprom2}
gives one for $1\le l<m\le 3$.
Furthermore,
\[
\lb{reprom3}
(x-1)\o^l-x\o^m+\o=x(\o^l-\o^m)+\o-\o^l
=(1-x)(\o^m-\o^l)+\o-\o^m.
\]
The second representation in \er{reprom3} gives
the second estimate in \er{estexpomx}
for $1\le l\le m\le 3$, and the third representation in \er{reprom3}
gives one for $1\le m<l\le 3$.~\BBox

\medskip

Introduce
the domain $\L_+$ in $\C_+$ by
$$
\L_+=\{\l\in\C_+:|\l|>1\},\qq \ol\L_+=\{\l\in\C:|\l|>1,\Im\l\ge 0\}.
$$
We proved in \cite[Lm~2.5]{BK25} that
if $\gu\in\cB(\ve)$,
then the following representation of the fundamental matrix
holds true:
\[
\lb{idXcX}
\F(x,\l)
=Z(\l) U W(x,\l)U^{-1}Z^{-1}(\l),\qq
\]
for all $(x,\l)\in[0,2]\ts\ol\L_+$,
where
\[
\lb{W1wtW1}
W(x,\l)=X(x,\l) W_1(x,\l)X^{-1}(0,\l),
\qq
W_1(x,\l)=e^{zx\O-{\O^2\/3z}\int_0^xp(s)ds},
\qq
W_1(1,\l)=e^{z\O},
\]
$Z,\O$, and $U$ are given by \er{4g.Om} and \er{defmaZ},
each matrix-valued function $X(\cdot,\l),\l\in\ol\L_+$,
is continuous on $[0,2]$,
each $X(x,\cdot),x\in[0,2]$,
is analytic on the domain $\L_+$, continuous on $\ol\L_+$, and satisfies
\[
\lb{estcX}
\det X=1,\qq
\sup_{x\in[0,2]}\max\{|X(x,\l)|,|X^{-1}(x,\l)|\}\le C,
\]
\[
\lb{ascX}
\sup_{x\in[0,2]}\max\Big\{|X(x,\l)-\1_3|,
|X^{-1}(x,\l)-\1_3|\Big\}\le {C\|\gu\|_1\/|z|^2},
\]
here and below we use the standard operator norm of matrices
 $|A|=\sup_{|h|=1}|Ah|,|h|^2=\sum_{j=1}^3|h_j|^2$.
Moreover,
each of the functions
$X(x,\l,\cdot),(x,\l)\in[0,2]\ts\ol\L_+$,
is analytic on the ball $\cB_\C(\ve)$.

The identities \er{simMt} and  \er{idXcX} imply
\[
\lb{repwtX}
\wt \F(x,\l) =J(\F^\top(x,\l))^{-1}J
=J Z^{-1}(\l) (U^{-1})^\top\wt W(x,\l)U^\top Z(\l) J,
\]
for all $(x,\l)\in[0,2]\ts\ol\L_+$, where $J$ is given by  \er{simMt},
\[
\lb{wtW1wtW1}
\wt W(x,\l)
=(W^\top(x,\l))^{-1}
=(X^\top(x,\l))^{-1}\wt W_1(x,\l)X^\top(0,\l),
\]
\[
\lb{defwtW0}
\wt W_1(x,\l)=W_1^{-1}(x,\l)=e^{-zx\O+{\O^2\/3z}\int_0^xp(s)ds},
\qq \wt W_1(1,\l)=e^{-z\O}.
\]
The definitions \er{W1wtW1} and \er{defwtW0}
and the  estimates \er{estexpom} imply
\[
\lb{pertW0}
\big|W_1(x,\l)\big|
\le C|e^{zx}|,\qq
\Big|W_1(x,\l)-e^{zx\O}
+{(e^{zx\O})''\/3z^3}\int_0^xp(s)ds\Big|
\le |e^{zx}|{C\|\gu\|^2\/|z|^2},
\]
\[
\lb{pertwtW0}
\big|\wt W_1(x,\l)\big|
\le C|e^{-\o zx}|,\qq
\Big|\wt W_1(x,\l)-e^{-zx\O}
-{(e^{-zx\O})''\/3z^3}\int_0^xp(s)ds\Big|
\le |e^{-\o zx}|{C\|\gu\|^2\/|z|^2},
\]
for all  $(x,\l)\in[0,2]\ts\ol\L_+$ and for some $C>0$.
We prove the following result.

\begin{lemma}
\lb{LmBest}
Let $(\l,\gu)\in\ol\L_+\ts\cB(\ve)$.
Then the matrix-valued functions $W,\F$ satisfy
\[
\lb{estX}
\sup_{x\in[0,2]}\max\big\{\big|W(x,\l)e^{-zx}\big|,
\big| \wt W(x,\l)e^{\o zx} \big|\big\}
\le C,
\]
\[
\lb{asX-X0}
\sup_{x\in[0,2]}\max
\big\{\big|\big(W(x,\l)-W_1(x,\l)\big) e^{-zx}\big|,
\big|\big(\wt W(x,\l)-\wt W_1(x,\l)\big)e^{\o zx}\big|\big\}
\le {C\|\gu\|_1\/|z|^2},
\]
\[
\lb{estPhi}
\sup_{x\in[0,2]}\max\big\{\big|\F_{jk}(x,\l)e^{-zx}\big|,
\big| \wt\F_{jk}(x,\l)e^{\o zx} \big|\big\}
\le {C\/|z|^{j-k}},
\]
\[
\lb{estfi-fi0}
\sup_{t\in[0,1]}\max
\big\{|\F_{jk}(1,t,\l)-\F_{0,jk}(1,\l)|e^{-\Re z},
|\wt\F_{jk}(1,t,\l)-\wt\F_{0,jk}(1,\l)|e^{-{1\/2}\Re z}\big\}\le
 {C \|\gu\|_1\/|z|^{2+j-k}},
\]
for all $j,k=1,2,3$ and
for some $C>0$.
\end{lemma}

\no {\bf Proof.} Let $(x,\l,\gu)\in[0,2]\ts\ol\L_+\ts\cB(\ve)$.
The definition \er{W1wtW1} and the estimates \er{estcX} imply
$$
|W(x,\l)|\le|X(x,\l)| |W_1(x,\l)||X^{-1}(0,\l)|
\le C |W_1(x,\l)|.
$$
The estimate \er{pertW0} gives the first estimate in \er{estX}.
Similarly, the definition \er{wtW1wtW1} and the estimates \er{estcX}
and \er{pertwtW0} yield the second estimate in \er{estX}.
The estimates \er{estX} and the identities \er{idXcX} and \er{repwtX}
give \er{estPhi}.

The definition \er{W1wtW1} implies
$$
\begin{aligned}
\big|W(x,\l)-W_1(x,\l)\big|
=\big|X(x,\l) W_1(x,\l)X^{-1}(0,\l)-W_1(x,\l)\big|
\\
\le\big|X(x,\l)-\1_3\big|
| W_1(x,\l)||X^{-1}(0,\l)|
+|W_1(x,\l)|\big|X^{-1}(0,\l)-\1_3\big|.
\end{aligned}
$$
Then the estimates \er{ascX} and \er{pertW0} imply the first estimate in
\er{asX-X0}. Similarly, the definition \er{wtW1wtW1} and the estimates
\er{ascX} and \er{pertwtW0} give the second estimate in
\er{asX-X0}.
The estimates \er{asX-X0} and
the identity $\int_0^1p(x+t)dx=0$ give \er{estfi-fi0}.~\BBox

\subsection{Traces of the monodromy matrix}
The characteristic determinant $D$ of  the monodromy matrix $M$,
given by \er{1c},
satisfies the standard identity
\[
\lb{cM}
D=(\t_1-\t)(\t_2-\t)(\t_3-\t)
=-\t^3+T\t^2-\wt T\t +1,
\]
where the entire functions
$T$ and $\wt T$ are given by
\[
\lb{defT}
T=\Tr M,\qqq \wt T=\Tr M^{-1}=\Tr \wt M,\qq \wt M=\wt \F(1,\cdot),
\]
$\wt \F$ is given by \er{deMc}.
In the unperturbed case $\gu=0$ we have
$\wt y_j^0(x,\l)= y_j^0(x,-\l)$, for all  $(x,\l)\in\R\ts\C$.
Traces $T_0$ and $\wt T_0$ of
the monodromy matrices $M_0=\Phi_0(1,\cdot)$ and
$\wt M_0=\wt \F_0(1,\cdot)$ satisfy
$T_0(\l)=3 y_1^0(1,\l)$, $\wt T_0(\l)=3\wt y_1^0(1,\l)=3 y_1^0(1,-\l)$.
We prove the following results.

\begin{lemma} Let $\gu\in\gH$. Then the functions $T$ and $\wt T$ satisfy
\[
\lb{estT}
|T(\l)|\le Ce^{\Re z},\qq |\wt T(\l)|\le Ce^{{1\/2}\Re z},\qq
|\dot T(\l)|\le {Ce^{\Re z}\/|z|^2},\qq
|\dot{\wt T}(\l)|\le {Ce^{{1\/2}\Re z}\/|z|^2},
\]
\[
\lb{estgrT}
\sup_{t\in[0,1]}\max\Big\{e^{-\Re z}\Big(\Big|{\pa T(\l,\gu)\/\pa p(t)}\Big|
+|z|\Big|{\pa T(\l,\gu)\/\pa q(t)}\Big|\Big),
e^{-{1\/2}\Re z}\Big(\Big|{\pa\wt T(\l,\gu)\/\pa p(t)}\Big|
+|z|\Big|{\pa\wt T(\l,\gu)\/\pa q(t)}\Big|\Big)\Big\}\le
{C\|\gu\|_1\/|z|^3},
\]
for all $\l\in\L_+$ and for some $C>0$,
\[
\lb{derTrT=0}
{\pa T(\l)\/\pa\gu(t)}\Big|_{\gu=0}=
{\pa \wt T(\l)\/\pa\gu(t)}\Big|_{\gu=0}=0,\qq \forall\qq
(t,\l)\in[0,2]\ts\C.
\]
\end{lemma}

\no {\bf Proof.}
 The identities \er{idXcX} and \er{repwtX} give
$$
T=\Tr\F(1)=\Tr W(1),\qq \wt T=\Tr\wt\F(1)=\Tr \wt W(1).
$$
here and below in this proof $T=T(\l),W(x)=W(1),\F(x)=\F(x,\l),...$
Then, the definitions \er{W1wtW1} and the estimates \er{estX} imply
the first two estimates in
\er{estT}. The identities \er{shX} and \er{derlfm}
and the definition \er{mtrH} yield
$$
\begin{aligned}
\dot T=\Tr\dot \F(1)=\int_0^1\Tr\big(\F(1)\F^{-1}(t) J_q\F(t)\big)dt
=\int_0^1\Tr\big(\F(1+t)\F^{-1}(t) J_q\big)dt
\\
=\int_0^1\Tr\big(\F(1,t)J_q\big)dt=\int_0^1\F_{13}(1,t)dt.
\end{aligned}
$$
The estimates \er{pertW0} and the identities \er{idXcX} give the third
estimate in \er{estT}.
The identity \er{symXM} implies $\wt T(\l,\gu)=T(-\l,\gu_*)$,
then
$$
\dot{\wt T}(\l,\gu)=-\dot T(-\l,\gu_*)
=-\int_0^1\F_{13}(1,t,-\l,\gu_*)dt
=-\int_0^1\wt\F_{13}(1,t,\l)dt.
$$
The relations \er{pertwtW0} and \er{repwtX}
give the fourth estimate in \er{estT}.

The identity \er{grX(1)} implies
\[
\lb{idderTrT}
\begin{aligned}
{\pa T\/\pa\gu(t)}=
\Tr {\pa \F(1)\/\pa\gu(t)}
=-\Tr\big(\F(1)\F^{-1}(t)( J_p, J_q)\F(t)\big)
=-\Tr\big(\F(t+1)\F^{-1}(t)( J_p, J_q)\big)
\\
=-\Tr\big(\F(1,t)( J_p, J_q)\big)
=-\big(\F_{12}(1,t)+\F_{23}(1,t),\F_{13}(1,t)\big),
\end{aligned}
\]
where we used the identity \er{shX} and the definitions \er{mtrH}.
The identity \er{idderTrT} gives
\[
\lb{grTat0pr}
{\pa T\/\pa\gu(t)}\Big|_{\gu=0}
=-\big(\F_{0,12}(1)+\F_{0,23}(1),\F_{0,13}(1)\big).
\]
Thus, ${\pa T\/\pa\gu(t)}|_{\gu=0}=\const$, therefore,
$\int_0^1{\pa T\/\pa\gu(t)}|_{\gu=0}\d\gu(t)dt=0$ for all $\d\gu\in\gH$.
The similar result for ${\pa \wt T\/\pa\gu(t)}|_{\gu=0}$ holds true,
which yields \er{derTrT=0}.

The identities \er{derTrT=0}, \er{idderTrT}, and \er{grTat0pr} give
$$
{\pa T\/\pa\gu(t)}
={\pa T\/\pa\gu(t)}-{\pa T\/\pa\gu(t)}\Big|_{\gu=0}
=-\big(\F_{12}(1,t)-\F_{0,12}(1)+\F_{23}(1,t)-\F_{0,23}(1),
\F_{13}(1,t)-\F_{0,13}(1)\big).
$$
The similar arguments and the identity \er{grwtX(1)} yield
$$
{\pa\wt T(\l)\/\pa\gu(t)}
=-\big(\wt\F_{12}(1,t)-\wt\F_{0,12}(1)+\wt\F_{23}(1,t)-\wt\F_{0,23}(1),
-\wt\F_{13}(1,t)+\wt\F_{0,13}(1)\big).
$$
The estimates \er{estfi-fi0} and the identities \er{idXcX}
and \er{repwtX} yield \er{estgrT}.~\BBox

\section{\lb{Sec3} The fundamental matrix at the three-point eigenvalues}
\setcounter{equation}{0}

\subsection{The unperturbed case}

Consider the unperturbed case $p=q=0$. The fundamental solutions
$ y_j$ of the unperturbed equation
$y'''=\l y$ have the form
\[
\lb{unpfs}
 y_1^0={e^{\o zx}+e^{\o^2 zx}+e^{zx}\/3},\qq
 y_2^0={\o^2e^{\o zx}+\o e^{\o^2 zx}+e^{zx}\/3z},\qq
 y_3^0={\o e^{\o zx}+\o^2e^{\o^2 zx}+e^{zx}\/3z^2},
\]
 and  the fundamental solutions $\wt y_j^0$
of the transpose unperturbed equation $-y'''=\l y$
satisfy
\[
\lb{wtyjtyj}
\wt y_j^0(x,\l)=(-1)^{j-1}y_j^0(-x,\l)
=y_j^0(x,-\l),\qq j=1,2,3,\qq (x,\l)\in\R\ts\C.
\]
Moreover,
\[
\lb{difwty0}
(y_j^0)'=y_{j-1}^0,\qq (\wt y_j^0)'=\wt y_{j-1}^0,\qq j=2,3.
\]
The eigenvalues $\m_n^0$ of the operator $\cL(0)$ and
the eigenvalues $\wt\m_n^0$ of the transpose operator $\wt\cL(0)$ have the form
$$
\iota_n:=\m_n^0=\wt\m_n^0=(2\nu n)^3,\qq \nu ={\pi \/\sqrt3}.
$$
Below we use the following notation for the solutions $y_j^0$ at the points
$\iota_n$:
\[
\lb{symyn0}
y_{j,n}^0(x)=y_j^0(x,\iota_n),\qq
\wt y_{j,n}^0(x)=\wt y_j^0(x,\iota_n)=y_{j,-n}^0(x),\qq n\in\Z_0,\qq j=1,2,3.
\]
We will often use the identities
\[
\lb{fremn}
2\o \nu n=- \nu n+i\pi n,\qq
2\o^2 \nu n=- \nu n-i\pi n.
\]

\begin{lemma} For all $n\in\N$ the fundamental solutions $ y_j^0$
of the unperturbed equation
$y'''=\l y$ and  the fundamental solutions $\wt y_j^0$
of the transpose unperturbed equation
$-y'''=\l y$ satisfy
\[
\lb{vp23tunp}
\begin{aligned}
 y_{1,n}^0={e^{2\nu nx}\/3}
\big(1+2e^{-3\nu nx}c_n(x)\big),
\\
 y_{2,n}^0={e^{2\nu nx}\/2\sqrt 3\pi n}
\big(1+a_n(x)e^{- 3\nu nx}\big),
\qq
 y_{3,n}^0={e^{2\nu nx}\/(2\pi n)^2}
\big(1+a_{-n}(x)e^{-3\nu nx}\big),
\end{aligned}
\]
\[
\lb{vp23tunp1}
 y_{2,n}^0(1)={\z_ne^{2\nu n}\/2\sqrt 3\pi n},
\qq
 y_{3,n}^0(1)={\z_ne^{2\nu n}\/(2\pi n)^2},
\]
\[
\lb{wtvp230}
\begin{aligned}
\wt y_{1,n}^0={e^{\nu nx}\/3}
\big(2c_n(x)+e^{-3\nu nx}\big),
\\
\wt y_{2,n}^0=-{e^{{\nu nx}}\/2\sqrt 3\pi n}
\big(a_{-n}(x)+e^{-3\nu nx}\big),
\qq
\wt y_{3,n}^0={e^{\nu nx}\/(2\pi n)^2}
\big(a_{n}(x)+e^{-3\nu nx}\big),
\end{aligned}
\]
\[
\lb{idvp2unp}
\wt y_{2,n}^0(1)={(-1)^ne^{\nu n}\z_n\/2\sqrt3\pi n},
\qq
\wt y_{3,n}^0(1)={(-1)^{n+1}e^{\nu n}\z_n\/(2\pi n)^2},
\]
\[
\lb{dotvp30mn}
\dot y_{3,n}^0(1)
={ e^{{2\nu n}}\z_n\/(2\pi n)^4}\Big(1
-{\sqrt3\/\pi n}\Big),
\qqq
\dot{\wt y_{3,n}^0}(1)
={(-1)^ne^{\nu n}\z_n\/(2\pi n)^4}\Big(1
+{\sqrt3\/\pi n}\Big),
\]
where
\[
\lb{anbn}
a_n(x)=\o e^{-i\pi nx }+\o^{2} e^{i\pi nx  }=\sqrt3s_n(x)-c_n(x),
\]
\[
\lb{defcnsn}
c_n(x)=\cos(\pi nx),\qq s_n(x)=\sin(\pi nx),
\]
\[
\lb{defzetan}
\z_n=1-(-1)^ne^{-3\nu n}.
\]
\end{lemma}

\no {\bf Proof.}
The identities \er{unpfs} and \er{fremn} give \er{vp23tunp}.
 Using $a_n(1)=(-1)^{n+1}$, we obtain \er{vp23tunp1}.
The identities \er{wtyjtyj}, \er{vp23tunp}, and
$a_n(-x)=a_{-n}(x)$  imply \er{wtvp230},
which yields \er{idvp2unp}.
The identities \er{wtyjtyj}, \er{vp23tunp1}, and
$\dot y_3^0(1,\l)={ 1\/3z^3}(y_2^0(1,\l)-2 y_3^0(1,\l))$,
give \er{dotvp30mn}.~\BBox

\subsection{Asymptotics of the fundamental matrix at the eigenvalues}
 Let $\gu\in\cB(\ve)$ and let $\l\in\{\m_n,\wt\m_n\},n\in\N$.
Then $\l$ is real and
\[
\lb{locmn}
|z-2\nu n|\le{C\|\gu\|_1\/n^2},\qq \nu={\pi\/\sqrt3},\qq z=\l^{1\/3},
\]
for some $C>0$, see \cite[Lm~3.8]{BK25}.
We prove the following estimates.

\begin{lemma}
Let $\gu\in\cB(\ve)$, let $\l\in\{\m_n,\wt\m_n\},n\in\N$,
and let $j,k=1,2,3$. Then the functions $W,\F$
given by \er{W1wtW1}, \er{deM},
respectively, satisfy
\[
\lb{asX-X0atmu0}
\sup_{x\in[0,2]}\max
\big\{e^{-2\nu nx}\big|W(x,\l)-W_1(x,\iota_n)\big|,
e^{-\nu n x}\big|\wt W(x,\l)-\wt W_1(x,\iota_n)\big|
\big\}
\le {C\|\gu\|_1\/n^2},
\]
\[
\lb{estXjkmu-mu0}
\begin{aligned}
\sup_{x\in[0,2]}\max\Big\{
e^{-{2\nu nx} }\Big|\F_{jk}(x,\l)-\F_{0,jk}(x,\iota_n)
+\rho_n(x)\F_{0,jk}''(x,\iota_n)\Big|,
\\
e^{-{\nu nx}}\Big|\wt \F_{jk}(x,\l)-\wt \F_{0,jk}(x,\iota_n)
-\rho_n(x)\wt\F_{0,jk}''(x,\iota_n)\Big|\Big\}
\le {C\|\gu\|_1\/n^{2+k-j}},\qq j,k=1,2,3,
\end{aligned}
\]
for some $C>0$, where $W_1$ and $\wt W_1$ are given by \er{W1wtW1} and \er{defwtW0},
\[
\lb{defXint}
\rho_n={1\/3(2\nu n)^3}\int_0^xp(s)ds.
\]
\end{lemma}

\no {\bf Proof.} Let $x\in[0,2]$. The estimates \er{asX-X0},  \er{estexpom},
and the definition \er{W1wtW1} give
$$
\big|W(x,\l)-W_1(x,\iota_n)\big|\le
\big|W(x,\l)-W_1(x,\l)\big|
+\big|W_1(x,\l)-W_1(x,\iota_n)\big|
\le {Ce^{2\nu nx}\|\gu\|_1\/n^{2}},
$$
for some $C>0$. These estimates and the similar estimates for
$\wt W$ yield \er{asX-X0atmu0}.
The estimates \er{asX-X0atmu0}, \er{pertW0}, and \er{pertwtW0}
and the identities \er{idXcX} and \er{repwtX}
give \er{estXjkmu-mu0}.~\BBox

\begin{lemma}

If $\gu\in\cB(\ve)$,
then
\[
\lb{estT-T0}
|T(\wt\m_n)-T_0(\iota_n)|\le  e^{2\nu n}{C\|\gu\|_1\/n^{2}},\qq
|\wt T(\wt\m_n)-\wt T_0(\iota_n)|\le   e^{\nu n}{C\|\gu\|_1\/n^{2}},
\]
\[
\lb{estTrwtTr}
\sup_{t\in[0,1]}\bigg(e^{-{2\nu n} }\Big(\Big|{\pa T(\wt\m_n)\/\pa p(t)}\Big|+
n\Big|{\pa T(\wt\m_n)\/\pa q(t)}\Big|\Big)
+
e^{-{\nu n} }\Big(\Big|{\pa\wt T(\wt\m_n)\/\pa p(t)}\Big|+
n\Big|{\pa\wt T(\wt\m_n)\/\pa q(t)}\Big|\Big)\bigg)
\le {C\|\gu\|_1\/n^{3}},
\]
for all $n\in\N$ and for some $C>0$.
\end{lemma}

\no {\bf Proof.}
The estimates \er{estXjkmu-mu0} give \er{estT-T0}.
The estimates \er{estgrT}  yield \er{estTrwtTr}.~\BBox

\section{\lb{Sec5} Eigenfunctions}
\setcounter{equation}{0}

\subsection{Derivatives of the fundamental solutions}
Let $(x,t,\l,\gu)\in\R\ts\R\ts\C\ts\gH$.
Introduce the function
\[
\lb{idcWvp}
\eta(x,t,\l)
=\det\ma y_1(x,\l)& y_2(x,\l)& y_3(x,\l)\\
 y_1(t,\l)& y_2(t,\l)& y_3(t,\l)\\
 y_1'(t,\l)& y_2'(t,\l)& y_3'(t,\l)\am.
\]
We prove the following formulas.

\begin{lemma}
Let $\gu\in\gH$. Then

i) The function $\eta$, given by \er{idcWvp}, satisfies
\[
\lb{idcWvp1}
\eta(x,t,\l)
= y_1(x,\l)\wt y_3(t,\l)-
 y_2(x,\l)\wt y_2(t,\l)+
 y_3(x,\l)\wt y_1(t,\l)= y_3(x-t,t,\l)
=\wt y_3(t-x,x,\l),
\]
for all $x,t\in\R$,
where $ y_3(x,t,\l)$ and $\wt y_3(x,t,\l)$ are given by \er{defvp3t}.

\no
ii) The fundamental solutions satisfy
\[
\lb{dervpj}
{\pa y_j(x,\l)\/\pa\gu(t)}
=\ca-\theta_j(x,t,\l)\1_{[0,x]}(t),x\in[0,1]
\\
-\theta_j(x,t,\l)-\theta_j(x,t+1,\l)\1_{[0,x-1]}(t),x\in[1,2]
\ac,
\]
for all $j=1,2,3,(x,t,\l)\in[0,2]\ts[0,1]\ts\C$,
where
\[
\lb{idvjvp}
\theta_j(x,t,\l)=\big(\{ y_j(t,\l),\eta(x,t,\l)\}_t;
 y_j(t,\l)\eta(x,t,\l)
\big),
\]
$$
\{ y_j(t,\l),\eta(x,t,\l)\}_t= y_j'(t,\l)\eta(x,t,\l)
- y_j(t,\l)\eta_t'(x,t,\l).
$$
\end{lemma}

\no {\bf Proof.}
i) The definition \er{idcWvp} and the identity \er{simMt} imply the first
identity in \er{idcWvp1}. Let $t\in\R$ be fixed.
The definition \er{idcWvp} shows that
$w(x)=\eta(x,t,\l)$ is a solution to Eq.~\er{1b}
satisfying the conditions
$$
w(t)=0,\qq  w'(t)=0,\qq w^{[2]}(t)=1.
$$
Then $w(x)= y_3(x,t)$, which yields the second identity in \er{idcWvp1}.

The first identity in \er{idcWvp1} and the identities \er{simMt} give
\[
\lb{idcWvptr}
\eta(x,t,\l)
=\det\ma\wt y_1(t,\l)&\wt y_2(t,\l)&\wt y_3(t,\l)\\
\wt y_1(x,\l)&\wt y_2(x,\l)&\wt y_3(x,\l)\\
\wt y_1'(x,\l)&\wt y_2'(x,\l)&\wt y_3'(x,\l)\am.
\]
Let $x$ be fixed.
Then the first identity in \er{idcWvp1} shows that
$\wt w(t)=\eta(x,t,\l)$ is a solution to Eq.~\er{1btr}.
The identity \er{idcWvptr} gives $\wt w(x)=0$,
$\wt w'(x)=0$, $\wt w^{[2]}(x)=1$.
Then $\wt w(x)=\wt y_3(t,x)$, which yields the third identity in \er{idcWvp1}.

ii) The definition \er{deM} and the identity \er{derpsifm} give
\[
\lb{Theta1j}
{\pa y_j(x)\/\pa\gu(t)}
={\pa \F_{1j}(x)\/\pa\gu(t)}
=\ca-\Theta_{1j}(x,t)\1_{[0,x]}(t),x\in[0,1]
\\
-\Theta_{1j}(x,t)-\Theta_{1j}(x,t+1)\1_{[0,x-1]}(t),x\in[1,2]\ac,
\]
where $\Theta(x,t,\l)=\Theta(x,t),...$
The definition \er{defcAxt} and the identities \er{shX} and \er{simMt}
imply
\[
\lb{thetathrphi}
\Theta(x,t)=\F(x-t,t)( J_p, J_q)\F(t)
=\F(x)\F^{-1}(t)( J_p, J_q)\F(t)=\F(x)J\wt \F^\top(t) J( J_p, J_q)\F(t).
\]
The definitions \er{mtrH} and \er{simMt} yield
$$
(\F J)_{1k}=(-1)^{k+1}\F_{1,4-k},\qq
\big(\wt\F^\top J J_q\F\big)_{kj}=\wt\F_{1k}\F_{1j},\qq
\big(\wt\F^\top J J_p\F\big)_{kj}
=\wt\F_{1k}\F_{2j}-\wt\F_{2k}\F_{1j}.
$$
Substituting these identities into \er{thetathrphi} and using
\er{deM} and \er{deMc}, we obtain
$$
\begin{aligned}
\Theta_{1j}(x,t)
=\sum_{k=1}^3 (-1)^{k+1}\F_{1,4-k}(x)
\big(\tilde \F_{1k}(t)\F_{2j}(t)-\tilde \F_{2k}(t)\F_{1j}(t),
\tilde \F_{1k}(t)\F_{1j}(t)\big)
\\
=\sum_{k=1}^3 (-1)^{k+1} y_{4-k}(x)
\big(\tilde  y_{k}(t) y_{j}'(t)-\tilde \vp'_{k}(t) y_{j}(t),
\tilde  y_{k}(t) y_{j}(t)\big)
\end{aligned}
$$
The identity \er{idcWvp1} and the definition  \er{idvjvp} imply
$$
\Theta_{1j}(x,t)
=\big(\eta(x,t) y_{j}'(t)-\eta_t'(x,t) y_{j}(t),\eta(x,t) y_{j}(t)\big)
=\theta_j(x,t).
$$
The identity \er{Theta1j} yields \er{dervpj}.~\BBox

\subsection{Functions $\vp,\p$}

Introduce the following solutions to  Eq.~\er{1b}:
\[
\lb{defef3p}
\varphi(x,\l)=\det\ma
0& y_2(x,\l)& y_3(x,\l)\\
 y_1(1,\l)& y_2(1,\l)& y_3(1,\l)\\
 y_1(2,\l)& y_2(2,\l)& y_3(2,\l)
\am,
\qq
\psi(x,\l)
=\det\ma y_2(x,\l)& y_3(x,\l)\\
 y_2(1,\l)& y_3(1,\l)\am,
\]
$(x,\l)\in\R\ts\C$.
The definition \er{defef3p} implies
\[
\lb{mg''0}
\varphi^{[2]}(0,\l)=\det\ma
 y_1(1,\l)& y_2(1,\l)\\
 y_1(2,\l)& y_2(2,\l)\am.
\]
If $\mu\in\C$ is an eigenvalue
of the operator $\cL$, then
 $\Delta(\mu)=0$, where $\Delta$ is given by \er{defsi}. This identity gives
\[
\lb{propcol}
 y_2(2,\mu)( y_2(1,\mu), y_3(1,\mu))
= y_2(1,\mu)( y_2(2,\mu), y_3(2,\mu)).
\]
The following lemma shows that in this case the functions $\varphi$
and  $\psi$, given by \er{defef3p},
are the corresponding eigenfunctions.

\begin{lemma} Let $\gu\in\gH$. Then

i) For all $\l\in\C$ the functions $\varphi$ and $\psi$ satisfy
\[
\lb{mg012}
\varphi(0,\l)=0,\qq \varphi(1,\l)=- y_1(1,\l)\Delta(\l),
\qq \varphi(2,\l)=- y_1(2,\l)\Delta(\l),
\]
\[
\lb{gh012}
\psi(0,\l)=\psi(1,\l)=0,\qq \psi(2,\l)=-\Delta(\l),
\]
\[
\lb{idght+1}
\psi(x+1,\l)
=- y_2(x,\l)\wt y_3(1,\l)- y_3(x,\l)\wt y_3'(1,\l),\qq x\in\R.
\]

ii) Let $\mu\in\C$ be an eigenvalue
of the operator $\cL$.
Then the functions $\varphi(x,\m)$
and $\psi(x,\m)$, given by \er{defef3p},
are eigenfunctions corresponding to the eigenvalue $\mu$.
Moreover,
\[
\lb{idef3p}
\varphi(x,\mu )=\det\ma y_1(x,\mu )& y_2(x,\mu  )& y_3(x,\mu  )\\
 y_1(1,\mu )& y_2(1,\mu  )& y_3(1,\mu )\\
 y_1(2,\mu  )& y_2(2,\mu )& y_3(2,\mu  )\am,\qq x\in\R.
\]
Let, in addition, $ y_2(1,\mu  )\ne 0$ and $ y_3(1,\mu  )\ne 0$. Then
$\varphi'(0,\mu  )\ne 0,\varphi^{[2]}(0,\mu  )\ne 0,\psi'(0,\mu  )\ne 0$
and the following identities hold:
\[
\lb{relgh}
\psi(x,\m)=-{ y_2(1,\m)\varphi(x,\m)\/\varphi^{[2]}(0,\m)},\qq x\in\R,
\]
\[
\lb{ef'1}
{\varphi'(1,\mu  )\/\varphi'(0,\mu  )}={\psi'(1,\mu)\/\psi'(0,\mu  )}=
-{\wt y_3(1,\mu )\/y_3(1,\mu  )}.
\]
Moreover, the functions $\theta_j,j=2,3$, given by \er{idvjvp}, satisfy
\[
\lb{thetaj2-1}
\theta_j(2,t,\mu )-{ y_2(2,\mu )\/ y_2(1,\mu )}\theta_j(1,t,\mu )
=-{\varphi^{[2]}(0,\mu )\/ y_2(1,\mu )}\big(\{ y_j(t,\mu ),\wt y_3(t,\mu )\};
 y_j(t,\mu )\wt y_3(t,\mu )\big),\qq t\in[0,1].
\]

\end{lemma}

\no {\bf Proof.}
i) The definitions \er{defef3p}
yield \er{mg012} and \er{gh012}.
The identity \er{shX} implies
$$
 y_j(x+1)= y_1(x) y_j(1)+ y_2(x) y_j'(1)+ y_3(x) y_j^{[2]}(1),
$$
for all $x\in\R,j=1,2,3$, here and below $\varphi(x)=\varphi(x,\l)$,
$ y_j(x)= y_j(x,\l),...$. Substituting these identities into \er{defef3p},
we obtain
$$
\psi(x+1)
=\det\ma y_1(x) y_2(1)+ y_2(x) y_2'(1)+ y_3(x) y_2^{[2]}(1)
& y_1(x) y_3(1)+ y_2(x) y_3'(1)+ y_3(x) y_3^{[2]}(1)\\
 y_2(1)& y_3(1)\am.
$$
The identity \er{simMt} gives
$\wt y_3=\det\ma y_2& y_3\\ y_2'& y_3'\am$ and
$\wt y_3'=\det\ma y_2& y_3\\ y_2^{[2]}& y_3^{[2]}\am$,
which yields \er{idght+1}.

ii) Let $\l=\mu$ be an eigenvalue
of the operator $\cL$.
Then $\Delta(\mu)=0$ and the definition \er{defef3p} gives \er{idef3p}.
The identities
\er{mg012} give
$\varphi(0)=\varphi(1)=\varphi(2)=0$.
The identities \er{gh012} yield $\psi(0)=\psi(1)=\psi(2)=0$.
Therefore, $\varphi$ and $\psi$ are eigenfunctions of the operator $\cL$
corresponding to the eigenvalue $\mu  $.

Assume that $\varphi'(0)=0$. Then $\varphi(x)=C y_3(x)$ for some $C\ne 0$,
which yields $ y_3(1)=0$. Thus, $ y_3(1)\ne 0$ implies $\varphi'(0)\ne 0$.
Similarly, $ y_3(1)\ne 0$ implies $\psi'(0)\ne 0$.
Moreover, the condition $\varphi^{[2]}(0)=0$ gives $\varphi(x)=C y_2(x)$
for some $C\ne 0$, which yields $ y_2(1)=0$.
Thus, $ y_2(1)\ne 0$ implies $\varphi^{[2]}(0)\ne 0$.

The identities \er{propcol} and \er{idef3p} yield
$$
\begin{aligned}
\varphi(x)=- y_1(1)\det\ma y_2(x)& y_3(x)\\
 y_2(2)& y_3(2)\am
+ y_1(2)
\det\ma
 y_2(x)& y_3(x)\\
 y_2(1)& y_3(1)\am
\\
=\Big( y_1(2)- y_1(1){ y_2(2)\/ y_2(1)}\Big)
\det\ma y_2(x)& y_3(x)\\
 y_2(1)& y_3(1)\am.
\end{aligned}
$$
The definition \er{defef3p} and the identity \er{mg''0} give \er{relgh}.

The identity \er{simMt} gives
$\det\ma   y_2(1)& y_3(1)\\ y_2'(1)&  y_3'(1)\am=\wt y_3(1)$.
The definition \er{defef3p} implies
\[
\lb{dh'01pr}
\psi'(1)
=\det\ma y_2'(1)& y_3'(1)\\
 y_2(1)& y_3(1)\am=-\wt y_3(1),\qq \psi'(0)= y_3(1).
\]
The identities
\er{relgh} and \er{dh'01pr} give \er{ef'1}.

Let $j=2,3, t\in[0,1]$.
The definition \er{idvjvp} implies
\[
\lb{theta2-1}
\theta_j(2,t)-{ y_2(2)\/ y_2(1)}\theta_j(1,t)
=\bigg(\Big\{ y_j(t),\eta(2,t)-{ y_2(2)\/ y_2(1)}\eta(1,t)\Big\};
 y_j(t)\Big(\eta(2,t)-{ y_2(2)\/ y_2(1)}\eta(1,t)\Big)\bigg).
\]
The first identity in \er{idcWvp1} and the identity $\Delta(\mu)=0$ provide
$$
\eta(2,t)-{ y_2(2)\/ y_2(1)}\eta(1,t)
=-{1\/ y_2(1)}\det\ma y_1(1)& y_2(1)\\ y_1(2)& y_2(2)\am\wt y_3(t)
=-{\varphi^{[2]}(0)\/ y_2(1)}\wt y_3(t),
$$
where we used \er{mg''0}. Substituting this identity into
\er{theta2-1}, we obtain \er{thetaj2-1}.~\BBox

\subsection{The eigenfunction $\varphi$}
The representation \er{idXcX} of the fundamental matrix $\F$ yields
the following representation for the eigenfunction $\varphi$ given by
\er{idef3p}.

\begin{lemma}
Let $\gu\in\cB(\ve)$.
If $\mu\in\C$ is an eigenvalue
of the operator $\cL$,
then the corresponding eigenfunction $\varphi$ satisfies
\[
\lb{mgthX}
\varphi^{[l-1]}(x,\mu)={i\xi_{l-1}(x,\mu)\/3\sqrt3\mu^{4-l\/3}},
\]
for all $l=1,2,3,x\in[0,2]$,
where
\[
\lb{defmX}
\xi_{l-1}(x,\l)=\det\ma
\varkappa_{l1}(x,\l)e^{zx\o-{\o^{2}\/3z}\int_0^xp(s)ds}
&\varkappa_{l2}(x,\l)e^{zx\o^2-{\o\/3z}\int_0^xp(s)ds}
&\varkappa_{l3}(x,\l)e^{zx-{1\/3z}\int_0^xp(s)ds}\\
\varkappa_{11}(1,\l)e^{z\o}&\varkappa_{12}(1,\l)e^{z\o^2}&\varkappa_{13}(1,\l)e^{z}\\
\varkappa_{11}(2,\l)e^{2z\o}&\varkappa_{12}(2,\l)e^{2z\o^2}&\varkappa_{13}(2,\l)e^{2z}\am,
\]
\[
\lb{defcXlj}
\varkappa_{lj}=\sum_{m=1}^3\o^{(l-1)m}X_{mj},
\]
the matrix-valued function $X$ satisfies \er{estcX} and \er{ascX}.
\end{lemma}

\no {\bf Proof.}
The representation \er{idXcX} of the fundamental matrix $\F$
may be rewritten in the form
\[
\lb{defcA}
\F(x,\l)=\Upsilon(x,\l)\Upsilon^{-1}(0,\l),\qq
\]
for all $(x,\l)\in[0,2]\ts\ol\L_+$, where
\[
\lb{defUps}
\Upsilon= Z UX e^{zx\O-{\O^2\/3z}\int_0^xp(s)ds},
\]
$Z$ and $U$ are defined by \er{4g.Om}
and \er{defmaZ}, respectively.
The fundamental matrix $\Upsilon$ has the form
\[
\lb{fsphi}
\Upsilon=\ma\phi_1&\phi_2&\phi_3\\
\phi_1'&\phi_2'&\phi_3'\\
\phi_1^{[2]}&\phi_2^{[2]}&\phi_3^{[2]}\am,
\]
where $\phi_1,\phi_2,\phi_3$ are some fundamental solutions to Eq.~\er{1b}.
The identities
$\det X=1$, $\det Z=\l$, $\det U=-i$, $\Tr\O=\Tr\O^2=0$,
and the definition \er{defUps} imply
\[
\lb{estdetcX}
\det \Upsilon=-i\l.
\]
Moreover, for all $x\in[0,2]$ we have
\[
\lb{fmrel}
\ma\phi_1(x,\l)&\phi_2(x,\l)&\phi_3(x,\l)\\
\phi_1(1,\l)&\phi_2(1,\l)&\phi_3(1,\l)\\
\phi_1(2,\l)&\phi_2(2,\l)&\phi_3(2,\l)\am
=\ma y_1(x,\l)& y_2(x,\l)& y_3(x,\l)\\
 y_1(1,\l)& y_2(1,\l)& y_3(1,\l)\\
 y_1(2,\l)& y_2(2,\l)& y_3(2,\l)\am
\Upsilon(0,\l).
\]
If $\mu$ is an eigenvalue
of the operator $\cL$, then
the identities \er{idef3p}, \er{estdetcX}, and \er{fmrel}
give
 \[
\lb{idef3pphi}
\varphi^{[l-1]}(x,\mu)={i\/\mu}
\det\ma\phi_1^{[l-1]}(x,\mu )&\phi_2^{[l-1]}(x,\mu )&\phi_3^{[l-1]}(x,\mu )\\
\phi_1(1,\mu )&\phi_2(1,\mu )&\phi_3(1,\mu )\\
\phi_1(2,\mu )&\phi_2(2,\mu )&\phi_3(2,\mu )\am,
\]
for all $l=1,2,3$.
The definitions \er{4g.Om}, \er{defmaZ}, \er{defUps}, and \er{fsphi}
yield
$$
\phi_j^{[l-1]}(x,\m)=\Upsilon_{lj}(x,\m)
={z^{l-1}\/\sqrt3}\sum_{m=1}^3\o^{(l-1)m} X_{mj}e^{zx\o^j-{\o^{2j}\/3z}\int_0^xp(s)ds}
={z^{l-1}\vk_{lj}\/\sqrt3}e^{zx\o^j-{\o^{2j}\/3z}\int_0^xp(s)ds},
$$
where $z=\m^{1\/3}$.
Substituting these identities into \er{idef3pphi}, we get \er{mgthX}.~\BBox

\medskip
Introduce the function
\[
\lb{defgb}
\chi(t,\l)=\ca \wt y_3(t,\l),0<t<1,\\
\wt y_3(t-2,\l)= y_3(2-t,t,\l),1<t<2\ac,
\]
here we used the last identity in \er{idcWvp1}.
Below we will use the following notations:
$$
\varphi_n(x)=\varphi(x,\m_n),\qq\psi_n(x)=\psi(x,\m_n),
\qq\chi_n(x)=\chi(x,\m_n),\qq y_{j,n}(x)=y_j(x,\m_n),\qq j=1,2,3,
$$
and the similar notations for the corresponding unperturbed functions
$\varphi_0=\varphi|_{\gu=0},\psi_0=\psi|_{\gu=0},\chi_0=\chi|_{\gu=0}$:
$$
\varphi_{0,n}=\varphi_n|_{\gu=0},\qq \psi_{0,n}=\psi_n|_{\gu=0},\qq
\chi_{0,n}=\chi_n|_{\gu=0},\qq y_{j,n}^0=y_{j,n}|_{\gu=0},\qq  j=1,2,3.
$$

\subsection{The unperturbed case}
In the unperturbed case  $\gu=0$ the function $\xi_{l-1}$, given by \er{defmX},
has the form
\[
\lb{defmX0}
\xi_{l-1}|_{\gu=0}={\xi_0^{(l-1)}\/z^{l-1}},\qq\text{where}\qq
\xi_0=\det\ma
e^{zx\o}&e^{zx\o^2}&e^{zx}\\
e^{z\o}&e^{z\o^2}&e^{z}\\
e^{2z\o}&e^{2z\o^2}&e^{2z}\am.
\]
Then the identity \er{mgthX} gives
\[
\lb{mgomX}
\varphi_{0,n}^{(l-1)}(x)=\varphi_0^{(l-1)}(x,\iota_n)
={i\xi_{l-1}|_{\gu=0}(x,\iota_n)\/3\sqrt3(2\nu n)^{4-l}}
={i\xi_0^{(l-1)}(x,\iota_n)\/(2\pi n)^3},
\]
for all $n\in\N$,
where $\iota_n=\mu_n^0=(2\nu n)^3$.
Introduce the functions
\[
\lb{defcAcBn}
w_{\varphi\chi,n}=\varphi_{0,n}''\chi_{0,n}'+\varphi_{0,n}'\chi_{0,n}'',
\qq
w_{\psi\chi,n}=\psi_{0,n}''\chi_{0,n}'+\psi_{0,n}'\chi_{0,n}'',
\qq n\in\N.
\]

\begin{lemma}
Let $n\in\N$.
The functions $\varphi_{0,n}$ and $\psi_{0,n}$
satisfy
\[
\lb{mgtmn}
\varphi_{0,n}(x)={\kappa_n(x)s_n(x)\/4(\pi n)^3},\qq
\varphi_{0,n}'(x)
={\kappa_n(x)b_{-n}(x)\/\sqrt3(2\pi n)^2},\qq
\varphi_{0,n}''(x)
=-{\kappa_n(x)b_{n}(x)\/6\pi n},
\]
\[
\lb{ghunp}
\psi_{0,n}(x)=(-1)^{n+1}e^{-\nu n}\varphi_{0,n}(x),
\]
for all $x\in\R$, where
\[
\lb{bn}
\kappa_n(x)=(-1)^{n+1}\z_ne^{(3-x)\nu n},\qq
b_n(x)=\sqrt3c_n(x)+s_n(x),
\]
$\z_n,c_n,s_n$ are given by \er{defzetan} and \er{defcnsn}.
Moreover, the functions $w_{\varphi\chi,n},w_{\psi\chi,n}$,
given by \er{defcAcBn}, satisfy
\[
\lb{cAn+cBn}
w_{\varphi\chi,n}(x)+w_{\psi\chi,n}(x+1)
={(-1)^{n}\z_n^2e^{3\nu n}a_{2n}(x)\/3(2\pi n)^2},\qq x\in[0,1].
\]

\end{lemma}

\no {\bf Proof.}
Let $n\in\N$.
The identities  \er{fremn} and the definition \er{defmX0} imply
\[
\lb{mX0atmu0}
\xi_0(x,\iota_n)=\det
\ma e^{(-\nu+i\pi) nx}
&e^{-(\nu+i\pi) nx}
&e^{2\nu nx}\\
(-1)^ne^{-\nu n}
&(-1)^ne^{-\nu n}
&e^{2\nu n}\\
e^{-2\nu n}
&e^{-2\nu n}
&e^{4\nu n}\am=i2(-1)^{n}\z_ne^{(3-x)\nu n}s_n(x).
\]
The identities \er{defmX0} and \er{mgomX}  give
$$
\varphi_{0,n}(x)={i\xi_0(x,\iota_n)\/(2\pi n)^3}.
$$
The identity \er{mX0atmu0} gives the first identity in \er{mgtmn},
which yields the second and third ones.

The identities \er{relgh} and  \er{mgtmn} imply
$$
\psi_{0,n}(x)=-{ y_{2,n}^0(1)\varphi_{0,n}(x)\/\varphi_{0,n}''(0)}
={2\sqrt3(-1)^{n+1}\pi n\/\z_n}e^{-3\nu  n} y_{2,n}^0(1)\varphi_{0,n}(x).
$$
Then the identity \er{vp23tunp1} gives \er{ghunp}.

We prove \er{cAn+cBn}. Let $x\in[0,1]$. Then the definition \er{defgb} implies
$\chi_{0,n}'=\wt y_{2,n}^0$, $\chi_{0,n}''=\wt y_{1,n}^0$.
The identities \er{wtvp230} yield
\[
\lb{gb0atx<1}
\chi_{0,n}'(x)=-{e^{{\nu nx}}\/2\sqrt 3\pi n}
\big(a_{-n}(x)+e^{-3\nu nx}\big),
\qq
\chi_{0,n}''(x)={e^{\nu nx}\/3}
\big(2c_n(x)+e^{-3\nu nx}\big),
\]
where $a_n,c_n$ are given by \er{anbn} and \er{defcnsn}.
Substituting the identities \er{mgtmn} and \er{gb0atx<1}
into \er{defcAcBn}, we obtain
$$
w_{\varphi\chi,n}(x)={\kappa_n(x)e^{\nu nx}\/3\sqrt 3(2\pi n)^2}\Big(b_{n}(x)
\big(a_{-n}(x)+e^{-3\nu nx}\big)
+b_{-n}(x)
\big(2c_n(x)+e^{-3\nu nx}\big)\Big).
$$
The identities $b_{n}a_{-n}+2c_n(x)b_{-n}=-\sqrt3a_{2n}$
and $b_n+b_{-n}=2\sqrt3c_n(x)$ give
\[
\lb{mgo'gb0''}
w_{\varphi\chi,n}(x)={(-1)^{n}\z_ne^{3\nu n}\/3(2\pi n)^2}
\big(a_{2n}(x)-2e^{-3\nu nx}c_n(x)\big).
\]

Let $x\in[1,2]$. Then the definition \er{defgb} and the identity
\er{wtyjtyj} give
$\chi_{0,n}'(x)=- y_{2,n}^0(2-x)$, $\chi_{0,n}''(x)= y_{1,n}^0(2-x)$.
The identities \er{vp23tunp} imply
\[
\lb{b0atx>1}
\chi_{0,n}'(x)=-{e^{2\nu n(2-x)}\/2\sqrt 3\pi n}
(1+a_{-n}(x)e^{- 3\nu n(2-x)}),
\qq
\chi_{0,n}''(x)={e^{2\nu n(2-x)}\/3}
(1+2e^{-3\nu n(2-x)}c_n(x)).
\]
Substituting the identities \er{ghunp}
 and \er{b0atx>1} into \er{defcAcBn} and using \er{mgtmn}, we obtain
$$
w_{\psi\chi,n}(x)={\z_n e^{3(2-x)\nu n}\/3\sqrt 3(2\pi n)^2}
\Big(b_{n}(x)\big(1+a_{-n}(x)e^{- 3\nu n(2-x)}\big)
+b_{-n}(x)\big(1+2e^{-3\nu n(2-x)}c_n(x)\big)\Big).
$$
The identities
$b_n+b_{-n}=2\sqrt3c_n(x)$ and
$b_na_{-n}+2b_{-n}c_n(x)=-\sqrt3a_{2n}$
yield
\[
\lb{gho''gbo'}
w_{\psi\chi,n}(x)={\z_n e^{3(2-x)\nu n}\/3(2\pi n)^2}
\big(2c_n(x)-a_{2n}(x)e^{- 3(2-x)\nu n}\big).
\]
The identities \er{mgo'gb0''} and \er{gho''gbo'} give \er{cAn+cBn}.~\BBox

\subsection{Asymptotics of the eigenfunctions}
We determine asymptotics of the eigenfunctions $\varphi$ and $\psi$,
the corresponding unperturbed functions satisfy \er{mgtmn} and \er{ghunp}.

\begin{lemma} Let $\gu\in\cB(\ve),n\in\N$. Then the functions
$\varphi_n,\psi_n$, and $\chi_n$ satisfy
\begin{multline}
\lb{asgb}
\max\Big\{\sup_{x\in[0,1]}
\Big(e^{-\nu nx}\Big|\chi_n^{[l]}(x)-\chi_{0,n}^{(l)}(x)
-\rho_n(x)\chi_{0,n}^{(l+2)}(x)\Big|\Big),
\\
\sup_{x\in[1,2]}
\Big(e^{2\nu n(x-2)}\Big|\chi_n^{[l]}(x)-\chi_{0,n}^{(l)}(x)
-\rho_n(x)\chi_{0,n}^{(l+2)}(x)\Big|\Big)\Big\}
\le {C\|\gu\|_1\/n^{4-l}},
\end{multline}
\[
\lb{asef3}
\sup_{x\in[0,2]}\Big(e^{(x-3)\nu n}\Big|\varphi_n^{[l]}(x)
-\varphi_{0,n}^{(l)}(x)
+\rho_n(x)\varphi_{0,n}^{(l+2)}(x)\Big|\Big)
\le {C\|\gu\|_1\/n^{5-l}},
\]
\[
\lb{asght+1}
\sup_{x\in[0,2]}\Big(e^{(x-2)\nu n}
\Big|\psi_n^{[l]}(x)-\psi_{0,n}^{(l)}(x)
+\rho_n(x)\psi_{0,n}^{(l+2)}(x)\Big|\Big)
\le{C\|\gu\|_1\/n^{5-l}},
\]
for all $l=0,1,2$ and
for some $C>0$, where $\rho_n$ is given by \er{defXint}.
\end{lemma}

\no {\bf Proof.}
We prove \er{asgb}. Let $n\in\N,l=0,1,2$.
If $x\in[0,1]$, then the definition \er{defgb} gives
$$
\chi_n^{[l]}(x)=\wt y_3^{[l]}(x,\m_n)=\wt \F_{l+1,3}(x,\m_n).
$$
The estimates \er{estXjkmu-mu0} yield \er{asgb} for $x\in[0,1]$.
Let $x\in[1,2]$. Then
$$
\chi_n^{[l]}(x,\gu)=\wt y_3^{[l]}(x-2,\m_n,\gu)
=(-1)^l y_3^{[l]}(2-x,\m_n,\gu^-)
=(-1)^l\F_{l+1,3}(2-x,\m_n,\gu^-),
$$
here we used the identity
$\wt y_3(x,\l,\gu)= y_3(-x,\l,\gu^-)$, which is obtained by
the substitution $-x$ instead of $x$ in Eq.~\er{1b}.
The estimates \er{estXjkmu-mu0} give
\[
\lb{asgb2-x}
\big|\chi_n^{[l]}(x,\gu)-(-1)^l\F_{0,l+1,3}(2-x,\iota_n)
+(-1)^l\rho_n(2-x,\gu^-)\F_{0,l+1,3}''(2-x,\iota_n)\big|
\le {C\|\gu\|_1\/n^{4-l}}e^{2\nu n(2-x) }.
\]
The definition \er{defXint} and the identities
$$
\int_0^{2-x}p(1-s)ds=\int_{x-1}^1p(t)dt=-\int_0^{x-1}p(t)dt
=-\int_0^xp(t)dt
$$
imply
$$
\rho_n(2-x,\gu^-)={1\/3(2\nu n)^3}\int_0^{2-x}p(1-s)ds
=-\rho_n(x,\gu).
$$
The estimate \er{asgb2-x} yields the second estimate in \er{asgb}.

We prove \er{asef3}. Let $n\in\N,l=0,1,2,x\in[0,2]$.
The definition \er{defcXlj} and the estimate \er{ascX} imply
$$
|\varkappa_{l+1,j}(x,\m_n)-\o^{lj}|\le{C\|\gu\|_1\/n^2},\qq j=1,2,3.
$$
The estimate \er{locmn} gives
$$
|e^{\m_n^{1\/3} a}-e^{2\nu na}|\le |e^{2\nu na}|{C\|\gu\|_1\/n^2}
$$
for all bounded $a\in\C$.
Moreover,
$$
\Big|e^{-{\o^j\/6\nu n}\int_0^xp(s)ds}-1+{\o^j\/6\nu n}\int_0^xp(s)ds\Big|
\le {C\|\gu\|_1\/n^2},\qq
j=1,2,3.
$$
Then the definition \er{defmX} and the identities \er{fremn} imply
$$
\xi_{l}(x,\m_n)
=\det\ma
\o^{l}e^{(-\nu+i\pi) nx}p_{n,2}(x)\b_{n,11}(x)
&\o^{2l}e^{-(\nu+i\pi) nx}p_{n,1}(x)\b_{n,12}(x)
&e^{2\nu nx}p_{n,3}(x)\b_{n,13}(x)\\
(-1)^ne^{-\nu n}\b_{n,21}(x)&(-1)^ne^{-\nu n}\b_{n,22}(x)&
e^{2\nu n}\b_{n,23}(x)\\
e^{-2\nu n}\b_{n,31}(x)&e^{-2\nu n}\b_{n,32}(x)&
e^{4\nu n}\b_{n,33}(x)\am,
$$
where
$$
p_{n,j}(x)=1-{\o^j\/6\nu n}\int_0^xp(s)ds,\qq
\b_{n,jk}(x)=1+\a_{n,jk}(x),\qq |\a_{n,jk}(x)|\le{C\|\gu\|_1\/n^2},\qq
j,k=1,2,3.
$$
Expanding the determinant along the third column yields
$$
\Bigg|\xi_{l}(x,\m_n)-\det\ma
\o^{l}e^{(-\nu+i\pi) nx}p_{n,2}(x)
&\o^{2l}e^{-(\nu+i\pi) nx}p_{n,1}(x)
&e^{2\nu nx}p_{n,3}(x)\\
(-1)^ne^{-\nu n}&(-1)^ne^{-\nu n}&
e^{2\nu n}\\
e^{-2\nu n}&e^{-2\nu n}&
e^{4\nu n}\am\Bigg|
\le e^{(3-x)\nu n}{C\|\gu\|_1\/n^2}.
$$
Then the identity \er{defmX0} implies
$$
\Big|\xi_{l}(x,\m_n)-{\xi_{0}^{(l)}(x,\iota_n)\/(2\nu n)^{l}}
+{\xi_{0}^{(l+2)}(x,\iota_n)\/3(2\nu n)^{l+3}}\int_0^xp(s)ds\Big|
\le e^{(3-x)\nu n}{C\|\gu\|_1\/n^2}.
$$
The identities \er{mgthX}
and \er{mgomX} imply \er{asef3}.

The estimates \er{estXjkmu-mu0}  and \er{asef3} give
$$
| y_{2,n}(1)- y_{2,n}^0(1)|\le e^{2\nu n}{C\|\gu\|_1\/n^3},
\qq
|\varphi_n^{[2]}(0)-\varphi_{0,n}''(0)|\le e^{3\nu n}{C\|\gu\|_1\/n^3},
$$
for all $n\in\N$ and
for some $C>0$. The identity \er{relgh}
yields \er{asght+1}.~\BBox

\section{\lb{Sec4} Gradient of the characteristic function $\Delta$}
\setcounter{equation}{0}

\subsection{Gradient of the function $\Delta$}
In the following lemma we calculate the gradient of
the characteristic function $\Delta$ given by \er{defsi}.

\begin{lemma}
\lb{Lmgragchf}
Let $\gu\in\gH$, let $\mu\in\C$ be an eigenvalue of the operator $\cL$,
and let $ y_2(1,\mu  )\ne 0$ and $ y_3(1,\mu  )\ne 0$.
Then the function $\Delta$ satisfies
\[
\lb{dermBev}
{\pa\Delta(\mu)\/\pa\gu(t)}
=\big(\{\varphi,\chi\};\varphi\chi\big)(t,\mu)
+\big(\{\psi,\chi\};\psi\chi\big)(t+1,\mu),\qq t\in[0,1],
\]
where $\{f,g\}=f'g-fg'$, $\varphi$, $\psi$, and $\chi$ are given by
\er{defef3p} and \er{defgb}.
\end{lemma}

\no {\bf Proof.}
Let $(t,\gu)\in[0,1]\ts\gH$ and let $\mu\in\C$ be an eigenvalue
of the operator $\cL$.
The definition \er{defsi} and the identity \er{propcol} give
\[
\lb{xiupr}
{\pa\Delta\/\pa\gu(t)}=\det
\ma{\pa y_2(1)\/\pa\gu(t)}&{\pa y_3(1)\/\pa\gu(t)}\\
 y_2(2)& y_3(2)\am
+\det\ma y_2(1)& y_3(1)\\
{\pa y_2(2)\/\pa\gu(t)}&{\pa y_3(2)\/\pa\gu(t)}\am
=\det
\ma
\a_2(t)&\a_3(t)\\
 y_2(1)& y_3(1)\am,
\]
where
\[
\lb{defab(t)}
\a_j(t)={ y_2(2)\/ y_2(1)}{\pa y_j(1)\/\pa\gu(t)}-{\pa y_j(2)\/\pa\gu(t)},\qq
j=2,3,
\]
here and below in this proof we write
$\Delta=\Delta(\mu), y_j(x)= y_j(x,\mu),...$
The identity \er{dervpj} implies
$$
{\pa y_j(1)\/\pa\gu(t)}=-\theta_j(1,t),\qq
{\pa y_j(2)\/\pa\gu(t)}=-\theta_j(2,t)-\theta_j(2,t+1),\qq j=2,3,
$$
where $\theta_j$ are given by \er{idvjvp}.
Substituting these identities into \er{defab(t)}, we obtain
$$
\a_j(t)=\theta_j(2,t)-{ y_2(2)\/ y_2(1)}\theta_j(1,t)+\theta_j(2,t+1),\qq
j=2,3.
$$
Substituting these identities into \er{xiupr}, we obtain
\[
\lb{gradxisum}
{\pa\Delta\/\pa\gu(t)}=A(t)+B(t),
\]
where
\[
\lb{defA(t)}
A(t)=\det
\ma\theta_2(2,t)-{ y_2(2)\/ y_2(1)}\theta_2(1,t)
&\theta_3(2,t)-{ y_2(2)\/ y_2(1)}\theta_3(1,t)\\
 y_2(1)& y_3(1)\am,
\]
\[
\lb{defB(t)}
B(t)=\det
\ma\theta_2(2,t+1)
&\theta_3(2,t+1)\\
 y_2(1)& y_3(1)\am.
\]
Substituting the identities \er{thetaj2-1} into \er{defA(t)} and using
the definition \er{defef3p} and the identity \er{relgh}, we obtain
\[
\lb{idA(t)}
\begin{aligned}
A(t)=-{\varphi^{[2]}(0)\/ y_2(1)}\det
\ma\big(\{ y_2(t),\wt y_3(t)\};
 y_2(t)\wt y_3(t)\big)
&\big(\{ y_3(t),\wt y_3(t)\};
 y_3(t)\wt y_3(t)\big)\\
 y_2(1)& y_3(1)\am
\\
=-{\varphi^{[2]}(0)\/ y_2(1)}\big(\{\psi(t),\wt y_3(t)\};\psi(t)\wt y_3(t)\big)
=\big(\{\varphi(t),\wt y_3(t)\};\varphi(t)\wt y_3(t)\big)
=\big(\{\varphi(t),\chi(t)\};\varphi(t)\chi(t)\big),
\end{aligned}
\]
here we used the definition \er{defgb}.
Substituting the definition \er{idvjvp} into \er{defB(t)}, we obtain
$$
B(t)
= \det\ma
\big(\{ y_2(s),\eta(2,s)\}; y_2(s)\eta(2,s)\big)&
\big(\{ y_3(s),\eta(2,s)\}; y_3(s)\eta(2,s)\big)\\
 y_2(1)& y_3(1)
\am,\qq s=t+1.
$$
The definition \er{defef3p} yields
\[
\lb{idB(t)}
B(t)=\big(\big\{\psi(s),\eta(2,s)\big\};
\psi(s)\eta(2,s)\big).
\]
The second identity in \er{idcWvp1} and the definition \er{defgb} imply
$\eta(2,s)=\wt y_3(s-2,2)=\wt y_3(s-2)=\chi(s)$, $s\in(1,2)$.
Substituting the identities \er{idA(t)} and \er{idB(t)} into
\er{gradxisum}, we obtain \er{dermBev}.~\BBox

\subsection{The unperturbed case}
In the unperturbed case $\gu=0$
the function $\Delta$, given by \er{defsi}, has the form
\[
\lb{mB0}
\Delta_0(\l)={8\/3\sqrt3 \l}\sin{\sqrt 3 z\/2}
\sin{\sqrt 3 \o z\/2} \sin{\sqrt 3 \o^2z\/2},
\]
see \cite{BK25}.
We prove the following results.

\begin{lemma}
Let $n\in\N$.
The functions $\dot\Delta_0$ and ${\pa\Delta\/\pa\gu}|_{\gu=0}$ satisfy
\[
\lb{dermBunp}
\dot\Delta_0(\iota_n)={(-1)^{n}\sqrt3e^{3\nu  n}\z_n^2\/(2\pi n)^5},
\]
\[
\lb{grmBunp}
{\pa\Delta(\iota_n)\/\pa\gu(t)}\Big|_{\gu=0}
={(-1)^{n+1}\sqrt3e^{3\nu  n}\z_n^2b_{2n}(t)\gv_n\/(2\pi n)^5},\qq t\in[0,1],
\]
where $\z_n$ is given by \er{defzetan},
$b_n$ is given by \er{bn}, and the vector $\gv_n$ has the form
\[
\lb{defgvn}
\gv_n={2\pi n\/3}\Big(1;-{\sqrt3\/2\pi n}\Big).
\]
\end{lemma}

\no {\bf Proof.} Let $n\in\N$.
The identity \er{mB0} implies
$$
\dot\Delta_0(\iota_n)={(-1)^{n}\sqrt3\/8(\pi n)^5}\sin\pi n \o \sin\pi n\o^2
={(-1)^{n}\sqrt3\/16(\pi n)^5}\big(\cosh(\sqrt3\pi n)-(-1)^n\big),
$$
which yields \er{dermBunp}.
We prove \er{grmBunp}. Let $t\in[0,1]$.
The identity \er{dermBev} and the definition \er{defgb} give
\[
\lb{grmBunppr}
{\pa\Delta(\iota_n)\/\pa\gu(t)}\Big|_{\gu=0}
=\big(\{\varphi_{0,n},\wt y_{3,n}^0\};\varphi_{0,n}\wt y_{3,n}^0\big)(t)
+\big(\big\{\psi_{0,n}(t+1),\wt y_{3,n}^0(t-1)\big\};
\psi_{0,n}(t+1) \wt y_{3,n}^0(t-1)\big),
\]
where $\{f,g\}=f'g-fg'$. The identities \er{mgtmn},
\er{ghunp}, and \er{bn} give
$$
\psi_{0,n}(t+1)=(-1)^{n+1}e^{-\nu n}\varphi_{0,n}(t+1)
=-e^{-2\nu n}\varphi_{0,n}(t).
$$
Substituting this identity into \er{grmBunppr}, we obtain
\[
\lb{gradDel1}
{\pa\Delta(\iota_n)\/\pa\gu(t)}\Big|_{\gu=0}
=\big(\{\varphi_{0,n}(t),r_n(t)\};
\varphi_{0,n}(t)r_n(t)\big),\qq
r_n(t)=\wt y_{3,n}^0(t)-e^{-2\nu n}\wt y_{3,n}^0(t-1).
\]
The definition \er{anbn} yields $a_{n}(t-1)=(-1)^na_n(t)$.
The identities \er{wtvp230} and the definition \er{defzetan}
 imply
\[
\lb{wty3-y3sh}
r_n(t)
={e^{\nu nt}\/(2\pi n)^2}\big(a_{n}(t)-a_{n}(t-1)e^{-3\nu n}\big)
={a_n(t)\z_ne^{\nu nt}\/(2\pi n)^2}.
\]
Moreover, the identities \er{difwty0} and \er{wtvp230}  give
\[
\lb{wty2-y2sh}
r_n'(t)
=\wt y_{2,n}^0(t)-e^{-2\nu n}\wt y_{2,n}^0(t-1)
=-{a_{-n}(t)\z_ne^{{\nu nt}}\/2\sqrt 3\pi n}.
\]
Substituting the identities \er{wty3-y3sh} and \er{wty2-y2sh}
into \er{gradDel1}, we obtain
$$
{\pa\Delta(\iota_n)\/\pa\gu(t)}\Big|_{\gu=0}
={\z_ne^{\nu nt}\/(2\pi n)^2}\Big(\varphi_{0,n}'(t)a_n(t)
+{2\pi n\/\sqrt 3}\varphi_{0,n}(t)a_{-n}(t);
\varphi_{0,n}(t)a_n(t)\Big).
$$
Then the identities \er{mgtmn} and \er{bn} give
\[
\lb{gradDel1}
{\pa\Delta(\iota_n)\/\pa\gu(t)}\Big|_{\gu=0}
={(-1)^{n+1}\z_n^2e^{3\nu n}\/\sqrt3(2\pi n)^4}\Big(b_{-n}(t)a_n(t)
+2s_n(t)a_{-n}(t);
{\sqrt 3\/\pi n}s_n(t)a_n(t)\Big).
\]
The definitions \er{anbn} and \er{bn} imply
$$
a_n(t)b_{-n}(t)+2a_{-n}(t)s_n(t)=b_{2n}(t)-2\sqrt3,
\qq
2a_n(t)s_n(t)=\sqrt3-b_{2n}(t).
$$
Substituting these identities into \er{gradDel1}, we obtain
$$
{\pa\Delta(\iota_n)\/\pa\gu(t)}\Big|_{\gu=0}
={(-1)^{n+1}\z_n^2e^{3\nu n}\/\sqrt3(2\pi n)^4}
\Big(b_{2n}(t)-2\sqrt3;
{\sqrt 3\/2\pi n}\big(\sqrt3-b_{2n}(t)\big)\Big).
$$
Using the identity $(\const,v)=0$ for all $v\in\gH$
we obtain \er{grmBunp}.~\BBox

\subsection{Asymptotics of derivatives of the function $\Delta$}
We determine asymptotics of the derivatives of the function $\Delta$,
the corresponding unperturbed functions satisfy \er{dermBunp} and \er{grmBunp}.

\begin{lemma}
Let $\gu\in\cB(\ve),n\in\N$. Then the function $\Delta$ satisfies
\[
\lb{asdotmB}
|\dot\Delta(\m_n)-\dot\Delta_0(\iota_n)|
\le {C\|\gu\|_1\/n^7}e^{3 \nu n},\qqq
|\dot\Delta(\m_n)|\ge {ce^{3 \nu n}\/n^5},
\]
\[
\lb{aspamBpsi}
\sup_{t\in[0,1]}\Big|{\pa\Delta(\m_n)\/\pa\gu(t)}
-{\pa\Delta(\iota_n)\/\pa\gu(t)}\big|_{\gu=0}\Big|
\le {C\|\gu\|_1\/n^6}e^{3\nu  n},
\]
for some $c,C>0$.
\end{lemma}

\no {\bf Proof.}
Let $\gu\in\cB(\ve)$ and let $n\in\N$. We have
\[
\lb{estdotxi1}
|\dot\Delta(\m_n)-\dot\Delta_0(\iota_n)|\le |\dot\Delta(\m_n)-\dot\Delta_0(\m_n)|
+|\dot\Delta_0(\m_n)-\dot\Delta_0(\iota_n)|.
\]
Cauchy's formula gives
$$
\dot\Delta(\m_n)-\dot\Delta_0(\m_n)={1\/3z_n^2}{d(\Delta(\l)-\Delta_0(\l))\/dz}\Big|_{\l=\m_n}
={1\/6\pi i z^2}\oint_{|\z-{2\nu n}|=1} {\Delta(\z^3)-\Delta_0(\z^3)\/(\z-z_n)^2}d\z,
$$
$z_n=\m_n^{1\/3}>0$.
This identity and  the estimate
\[
\lb{estwtmB}
\Big|{\Delta(\l)\/\Delta_0(\l)}-1\Big|\le {C\|\gu\|\/|z|^2},\qq
\l\in\ol\L_+\sm\cup_{n\in\Z}\cD_{n},
\]
from \cite[Lm~3.6]{BK25}
imply
\[
\lb{estdotmBpr}
|\dot\Delta(\m_n)-\dot\Delta_0(\m_n)|
\le{C\/n^2}\max_{\l\in\pa\cD_n}|\Delta(\l)-\Delta_0(\l)|
\le{C\|\gu\|_1\/n^4}\max_{\l\in\pa\cD_n}|\Delta_0(\l)|,
\]
for some $C>0$, where the domains $\cD_{n}$ are given by \er{DomcD}.
The identity \er{mB0} yields
\[
\lb{estxi0}
|\Delta_0(\l)|\le{8\/3\sqrt3 |\l|}
e^{{\sqrt 3 \/2}(|\Im z|+ |\Im\o z|+|\Im \o^2z|)}={8\/3\sqrt3 |\l|}
e^{{3 \/2}(\Re z+{|\Im z|\/\sqrt3})},
\qq \l\in\C.
\]
Then $\max_{\l\in\pa\cD_n}|\Delta_0(\l)|\le{C\/n^3}e^{3\nu  n}$
and the estimate \er{estdotmBpr} gives
\[
\lb{estdotmB}
|\dot\Delta(\m_n)-\dot\Delta_0(\m_n)|
\le{C\|\gu\|_1e^{3\nu  n}\/n^7},
\]
for some $C>0$.
The estimate \er{locmn} gives
$|z_n-{2\nu n}|\le{C\|\gu\|_1\/n^2}$.
Moreover,
$$
\begin{aligned}
\dot\Delta_0(\m_n)-\dot\Delta_0(\iota_n)
={1\/3z_n^2}{d\Delta_0(\l)\/dz}\Big|_{\l=\m_n}
-{1\/3({2\nu n})^2}{d\Delta_0(\l)\/dz}\Big|_{\l=\iota_n}
\\
={1\/6\pi i}\oint_{|\z-{2\nu n}|=1}\Big({1\/z_n^2}{1\/(\z-z_n)^2}
-{1\/({2\nu n})^2}{1\/(\z-{2\nu n})^2}\Big)\Delta_0(\z^3)d\z.
\end{aligned}
$$
Using \er{estxi0} and the estimate
$$
\Big|{1\/z_n^2}{1\/(\z-z_n)^2}
-{1\/({2\nu n})^2}{1\/(\z-{2\nu n})^2}\Big|
\le \max_{z\in[2\nu n,z_n]}\Big|{d\/dz}\Big({1\/z^2(\z-z)^2}\Big)\Big|
|z_n-{2\nu n}|\le{C\|\gu\|_1\/n^4},
$$
for all $\z:|\z-{2\nu n}|=1$, and for some $C>0$,
we obtain
\[
\lb{estdotxi2}
|\dot\Delta_0(\m_n)-\dot\Delta_0(\iota_n)|
\le {C\|\gu\|_1\/n^7}e^{ 3 \nu n},
\]
for some $C>0$.
Substituting the estimates \er{estdotmB} and \er{estdotxi2} into \er{estdotxi1}
we obtain the first estimate in \er{asdotmB}.
Moreover, the identity \er{dermBunp}, the definition
\er{defzetan}, and  the first estimate in \er{asdotmB}
give
$$
|\dot\Delta(\m_n)|\ge|\dot\Delta_0(\iota_n)|-|\dot\Delta(\m_n)-\dot\Delta_0(\iota_n)|
\ge\Big({\sqrt3\z_n^2\/(2\pi)^5}-{C\|\gu\|_1\/n^2}\Big){e^{3 \nu n}\/n^5}.
$$
The estimate $\z_n\ge 1-e^{-2\sqrt3\pi}$ for all $n\in\N$ yields the second estimate in \er{asdotmB}.

We prove  \er{aspamBpsi}.
Let $n\in\N$. The identity \er{dermBev} implies
\[
\lb{pamnpaq1}
{\pa\Delta(\mu_n)\/\pa q(t)}
=\varphi_n(t)\chi_n(t)
+\psi_n(t+1)\chi_n(t+1),\qq t\in[0,1].
\]
The estimates \er{estPhi} imply
$$
\sup_{t\in[0,1]}\big(e^{-\nu nt}|\wt y_3(t,\m_n)|\big)
=\sup_{t\in[0,1]}\big(e^{-\nu nt}|\wt \F_{13}(t,\m_n)|\big)
\le {C\/n^2},
$$
$$
\sup_{t\in[1,2]}\big(e^{{2\nu n(t-2)}}| y_3(2-t,t,\m_n)|\big)
=\sup_{t\in[1,2]}\big(e^{{2\nu n(t-2)}}|\F_{13}(2-t,t,\m_n)|\big)
\le {C\/n^2},
$$
and then the definition \er{defgb} gives
\[
\lb{est101}
\max\Big\{\sup_{t\in[0,1]}\big(e^{-\nu nt}|\chi_n(t)|\big)
,\sup_{t\in[1,2]}\big(e^{{2\nu n(t-2)}}|\chi_n(t)|\big)
\Big\}
\le {C\/n^2},
\]
for some $C>0$.
The estimates \er{asgb}, \er{asef3}, and \er{asght+1}
together with the definitions
\er{defXint}, \er{defgb}, and \er{ghunp} and the identity \er{mgtmn}
give
\[
\lb{est102}
\sup_{t\in[0,2]}\big(e^{(t-3)\nu n}|\varphi_{0,n}(t)|\big)
\le {C\/n^{3}},\qq
\sup_{t\in[0,2]}\Big(e^{(t-3)\nu n}\Big|\varphi_n(t)-\varphi_{0,n}(t)\Big|\Big)
\le {C\|\gu\|_1\/n^{4}},
\]
\[
\lb{est103}
\sup_{t\in[0,2]}\big(e^{(t-2)\nu n}|\psi_{0,n}(t)|\big)
\le {C\/n^{3}},
\qq
\sup_{t\in[0,2]}\Big(e^{(t-2)\nu n}
\Big|\psi_n(t)-\psi_{0,n}(t)\Big|\Big)
\le{C\|\gu\|_1\/n^{4}},
\]
\[
\lb{est104}
\sup_{t\in[0,1]}
\Big(e^{-\nu nt}\Big|\chi_n(t)-\chi_{0,n}(t)\Big|\Big)
\le {C\|\gu\|_1\/n^{3}},\qq
\sup_{t\in[1,2]}
\Big(e^{2\nu n(t-2)}\Big|\chi_n(t)-\chi_{0,n}(t)\Big|\Big)\Big\}
\le {C\|\gu\|_1\/n^{3}}.
\]
The estimates \er{est101}, \er{est102}, and \er{est104}
imply
$$
\begin{aligned}
\sup_{t\in[0,1]}\Big|\varphi_n(t)\chi_n(t)-\varphi_{0,n}(t)\chi_{0,n}(t)\Big|\le
\sup_{t\in[0,1]}\Big(
|\varphi_n(t)-\varphi_{0,n}(t)||\chi_n(t)|
+|\varphi_{0,n}(t)||\chi_n(t)-\chi_{0,n}(t)|
\Big)
\\
\le e^{3\nu  n}{C\|\gu\|_1\/n^6},
\end{aligned}
$$
for some $C>0$. The estimates \er{est101},  \er{est103}, and \er{est104}
provide
$$
\begin{aligned}
\sup_{t\in[1,2]}\Big(e^{3\nu  n(t-2)}
|\psi_n(t)\chi_n(t)-\psi_{0,n}(t)\chi_{0,n}(t)|\Big)
\\
\le\sup_{t\in[1,2]}\Big(e^{3\nu  n(t-2)}
\big(|\psi_n(t)-\psi_{0,n}(t)||\chi_n(t)|
+|\psi_{0,n}(t)||\chi_n(t)-\chi_{0,n}(t)|\big)\Big)
\le {C\|\gu\|_1\/n^6},
\end{aligned}
$$
for some $C>0$.
Substituting these estimates into the identity \er{pamnpaq1}
we obtain
\[
\lb{aspamBq}
\sup_{t\in[0,1]}\bigg|{\pa\Delta(\m_n)\/\pa q(t)}
-{\pa\Delta(\iota_n)\/\pa q(t)}\Big|_{\gu=0}\bigg|
\le e^{3\nu  n}{C\|\gu\|_1\/n^6},
\]
for some $C>0$.

The identity \er{dermBev} implies
\[
\lb{pamupap1}
{\pa\Delta(\mu_n)\/\pa p(t)}
=\{\varphi_n,\chi_n\}(t)+\{\psi_n,\chi_n\}(t+1),\qq t\in[0,1],
\]
where $\{f,g\}=f'g-fg'$.
If  $ t\in[0,1]$, then the estimates \er{asgb}, \er{asef3} and the identities
$\varphi_0'''=\l\varphi_0$, $\chi_0'''=-\l\chi_0$
give
$$
\begin{aligned}
\{\varphi,\chi\}
=\Big(\varphi_0'-\rho_n\varphi_0'''+e^{\nu n (3-t)}{\a_{1,n}\/n^{4}}\Big)
\Big(\chi_0+\rho_n\chi_0''+e^{\nu n t}{\a_{2,n}\/n^{4}}\Big)
\\
-\Big(\varphi_0-\rho_n\varphi_0''+e^{\nu n (3-t)}{\a_{3,n}\/n^{5}}\Big)
\Big(\chi_0'+\rho_n\chi_0'''+e^{\nu n t}{\a_{4,n}\/n^{3}}\Big)
=\{\varphi_0,\chi_0\}
+\rho_nw_{\varphi\chi,n}
+e^{3\nu n}{\a_n\/n^{6}},
\end{aligned}
$$
where
$|\a_{j,n}(t)|\le C\|\gu\|_1,|\a_n(t)|\le C\|\gu\|_1$,
$\rho_n$ is given by \er{defXint},
$w_{\varphi\chi,n}$ is given by \er{defcAcBn},
here and below we write $\varphi=\varphi_n,\varphi_0=\varphi_{0,n},...$
If  $ t\in[1,2]$, then the estimates \er{asgb}, \er{asght+1} and
the identities  $\psi_0'''=\l\psi_0$, $\chi_0'''=-\l\chi_0$
give
$$
\begin{aligned}
\{\psi,\chi\}
=\Big(\psi_0'-\rho_n\psi_0'''+e^{\nu n (2-t)}{\b_{1,n}\/n^{4}}\Big)
\Big(\chi_0+\rho_n\chi_0''+e^{2\nu n (2-t)}{\b_{2,n}\/n^{4}}\Big)
\\
-\Big(\psi_0-\rho_n\psi_0''+e^{\nu n (2-t)}{\b_{3,n}\/n^{5}}\Big)
\Big(\chi_0'+\rho_n\chi_0'''+e^{2\nu n (2-t)}{\b_{4,n}\/n^{3}}\Big)
=\{\psi_0,\chi_0\}
+\rho_nw_{\psi\chi,n}
+e^{3\nu n(2-t)}{\b_n\/n^{6}},
\end{aligned}
$$
where
$|\b_{j,n}(t)|\le C\|\gu\|_1,|\b_n(t)|\le C\|\gu\|_1$,
$w_{\psi\chi,n}$ is given by \er{defcAcBn}.
The identities \er{pamupap1} and \er{cAn+cBn} give
$$
{\pa\Delta(\m_n)\/\pa p(t)}
={\pa\Delta(\iota_n)\/\pa p(t)}\Big|_{\gu=0}
+{(-1)^{n}\z_n^2e^{3\nu n}\/3(2\pi n)^2}a_{2n}(t)\rho_n(t)
+e^{3\nu n}{\g_n(t)\/n^{6}},
$$
where $|\g_n(t)|\le C\|\gu\|_1$,
$a_{2n}$ and $\rho_n$ are given by \er{anbn} and \er{defXint},
respectively.

If $\d p\in\cH_1$, then integrating by parts, we obtain
$$
\Big|\int_0^1{\pa\Delta(\mu_n)\/\pa p(t)}\d p(t)-
\int_0^1{\pa\Delta(\iota_n)\/\pa p(t)}\Big|_{\gu=0}\d p(t)\Big|
\le e^{3\nu  n}{C\|\gu\|_1\/n^6}.
$$
This yields
$$
\sup_{t\in[0,1]}\bigg|{\pa\Delta(\m_n)\/\pa p(t)}
-{\pa\Delta(\iota_n)\/\pa p(t)}\Big|_{\gu=0}\bigg|
\le e^{3\nu  n}{C\|\gu\|_1\/n^6},
$$
for some $C>0$.
This estimate and  the estimate \er{aspamBq} give \er{aspamBpsi}.~\BBox

\section{\lb{Sec6} Gradients of the three-point eigenvalues}
\setcounter{equation}{0}

\subsection{Analyticity}
In the following lemma we prove that for each $n\in\Z_0$
the function $\mu_n$ is analytic on the ball $\cB_\C(\ve)$.
Recall that $\ve>0$ is some absolute small constant
that does not depend on anything.

\begin{lemma}
\lb{Lm3-pev}

Each of the functions $\mu_n,n\in\Z_0$,
is analytic on $\cB_\C(\ve)$. Moreover,
\[
\lb{idgramu}
{\pa\mu_n(\gu)\/\pa\gu}
=-{{\pa \/\pa \gu}\Delta(\l,\gu)\/\dot\Delta(\l,\gu)}\Big|_{\l=\mu_n(\gu)}.
\]

\end{lemma}

\no {\bf Proof.}
 Let $n\in\Z_0$ and let $\gu\in\cB_\C(\ve)$.
Then the function $\Delta(\cdot,\gu)$ has exactly one simple zero $\mu_n(\gu)$
in the domain $\cD_n$ and $\Delta(\l,\gu)\ne 0$, as $\l\in\pa\cD_n$.
This yields
$$
\mu_n(\gu)={1\/2\pi i}\oint_{\pa\cD_n}\l{\dot\Delta(\l,\gu)\/\Delta(\l,\gu)}d\l
=-{1\/2\pi i}\oint_{\pa\cD_n}\log\Delta(\l,\gu)d\l.
$$
Each of the functions $\Delta(\l,\cdot),\dot\Delta(\l,\cdot),\l\in\C$,
is analytic with respect to $\gu\in\gH_\C$.
Then each $\mu_n,n\in\Z_0$, is analytic on $\cB_\C(\ve)$ and
$$
{\pa\mu_n(\gu)\/\pa\gu}=-{1\/2\pi i}{\pa \/\pa\gu}\oint_{\pa\cD_n}\log\Delta(\l,\gu)d\l
=-{1\/2\pi i}\oint_{\pa\cD_n}{{\pa \/\pa \gu}\Delta(\l,\gu)\/\Delta(\l,\gu)}d\l,
$$
which yields \er{idgramu}.~\BBox

\subsection{The unperturbed case}
We prove the following results for the unperturbed case $p=q=0$.

\begin{proposition}
Let $n\in\N$.
Then the functions $\mu_n,\wt\mu_n$ satisfy
\[
\lb{grmuunp}
{\pa\m_n\/\pa\gu(t)}\Big|_{\gu=0}
=b_{2n}(t)\gv_n,
\qq
{\pa \wt\m_n\/\pa \gu(t)}\Big|_{\gu=0}
=b_{-2n}(t)\gv_n,
\]
where $b_n$ and $\gv_n$ are given by \er{bn} and \er{defgvn},
respectively.

\end{proposition}

\no {\bf Proof.} Let $n\in\N$.
Substituting the identities \er{dermBunp} and \er{grmBunp}
into \er{idgramu}, we obtain the first identity in \er{grmuunp}.
We prove  the second one.
Let $\gu\in\gH, n\in\N$. Then we have
$$
d_\gu\m_n|_{\gu=0}(\d\gu)
=\int_0^1\Big({\pa\m_n(\gu)\/\pa\gu(x)}\Big|_{\gu=0},\d\gu(x)\Big)dx
=\int_0^1\Big({\pa\m_n(\gu)\/\pa p(x)}\Big|_{\gu=0}\d p(x)
+{\pa\m_n(\gu)\/\pa q(x)}\Big|_{\gu=0}\d q(x)\Big)dx,
$$
for all $\d\gu\in\gH$.
The identity \er{grmuunp} and the definitions \er{defgvn}, \er{bn}
give
\[
\lb{difmunpsiunp}
d_\gu\m_n|_{\gu=0}(\d\gu)={2\pi n\/3}\Big(
\sqrt3\wh{\d p_{cn}}+\wh{\d p_{sn}}
-{\sqrt3\/2\pi n}
\big(\sqrt3\wh{\d q_{cn}}+\wh{\d q_{sn}}\big)\Big).
\]
The identities \er{symev} and  $\wt\m_n(\gu)=-\m_{-n}(\gu_*)$ imply
$\wt\m_n(\gu)=\m_{n}(\gu^-)$.
This yields
$$
\begin{aligned}
d_\gu\wt\m_n|_{\gu=0}(\d\gu)
=\int_0^1\Big({\pa \wt\m_n(\gu)\/\pa\gu(x)}\Big|_{\gu=0},\d\gu(x)\Big)dx
=\int_0^1\Big({\pa \m_{n}(\gu^-)\/\pa\gu(x)}\Big|_{\gu=0},\d\gu(x)\Big)dx
\\
=\int_0^1\Big({\pa \m_{n}(\gu^-)\/\pa\gu(1-x)}\Big|_{\gu=0},\d\gu(1-x)\Big)dx
=\int_0^1\Big({\pa \m_{n}(\gu^-)\/\pa\gu^-(x)}\Big|_{\gu^-=0},\d\gu^-(x)\Big)dx
=d_{\gu^-}\m_n|_{\gu^-=0}(\d\gu^-).
\end{aligned}
$$
By replacing $\gu$ with $\gu^-$ in the identity \er{difmunpsiunp}
we obtain
$$
d_\gu\wt\m_n|_{\gu=0}(\d\gu)
=d_{\gu^-}\m_n|_{\gu^-=0}(\d\gu^-)={2\pi n\/3}\Big(
\sqrt3\wh{\d p_{cn}}-\wh{\d p_{sn}}
-{\sqrt3\/2\pi n}
\big(\sqrt3\wh{\d q_{cn}}-\wh{\d q_{sn}}\big)\Big),
$$
which yields the second identity in \er{grmuunp}.~\BBox

\subsection{Asymptotics of the derivative of the eigenvalue}

We determine asymptotics of the Frech\'et derivatives of the eigenvalues
of the operators $\cL$ and $\wt\cL$.

\medskip

\no {\bf Proof of Theorem~\ref{Th3pram}.}
The estimate \er{as3pev} is proved in \cite[Th~1.1]{BK25}.
We prove \er{aspamun}.
Let $\gu\in\cB(\ve),n\in\N$. For  $f=p$ or $f=q$
the identity \er{idgramu} gives
\[
\lb{estdermu-mu0pr}
\begin{aligned}
\bigg|{\pa\mu_n\/\pa f}-{\pa\mu_n\/\pa f}\big|_{\gu=0}\bigg|
=\bigg|{{\pa \/\pa f}\Delta(\mu_n)\/\dot\Delta(\m_n)}
-{{\pa \/\pa f}\Delta(\iota_n)|_{\gu=0}\/\dot\Delta_0(\iota_n)}\bigg|
\\
\le\Big|{\pa\Delta(\mu_n)\/\pa f}\Big|\bigg|{1\/\dot\Delta(\m_n)}
-{1\/\dot\Delta_0(\iota_n)}\bigg|
+\bigg|{\pa \Delta(\mu_n)\/\pa f}
-{\pa \Delta(\iota_n)\/\pa f}\Big|_{\gu=0}\bigg|
{1\/|\dot\Delta_0(\iota_n)|}.
\end{aligned}
\]
The estimates \er{asdotmB} imply
$$
\Big|{1\/\dot\Delta(\m_n)}
-{1\/\dot\Delta_0(\iota_n)}\Big|
={|\dot\Delta(\m_n)-\dot\Delta_0(\iota_n)|\/|\dot\Delta(\m_n)||\dot\Delta_0(\iota_n)|}
\le Cn^3e^{-3 \nu n}\|\gu\|_1,
$$
for some $C>0$.
The identity \er{grmBunp} and the estimates \er{aspamBpsi}
yield
$$
\sup_{t\in[0,1]}\Big|{\pa\Delta(\m_n)\/\pa \gu(t)}\Big|
\le {C\|\gu\|_1\/n^4}e^{3\nu  n}.
$$
Substituting these estimates and the relations \er{dermBunp}
and \er{aspamBpsi} into \er{estdermu-mu0pr}
we obtain
$$
\sup_{t\in[0,1]}
\Big|{\pa\m_n\/\pa \gu(t)}-{\pa\m_n\/\pa \gu(t)}\big|_{\gu=0}\Big|
\le {C\|\gu\|_1\/n},
$$
for some $C>0$. Then the identity \er{grmuunp} gives \er{aspamun}
for $n>0$.

We prove the estimate for $n<0$.
Let $n\in\N$ and let $\d\gu=(\d p,\d q)\in\gH$. The symmetry \er{symev} gives
$$
\m_{-n}(\gu+\d\gu)=-\m_{n}(\gu_*^-+\d\gu_*^-)
=-\m_{n}(\gu_*^-)-d_{\gu_*^-}\m_{n}(\d\gu_*^-)+o(\d\gu)
=\m_{-n}(\gu)-d_{\gu_*^-}\m_{n}(\d\gu_*^-)+o(\d\gu).
$$
Moreover,
$$
\m_{-n}(\gu+\d\gu)=\m_{-n}(\gu)+d_\gu \m_{-n}(\d\gu)+o(\d\gu).
$$
Comparing these asymptotics, we get
\[
\lb{dm-ndm*n}
d_\gu \m_{-n}(\d\gu)=-d_{\gu_*^-}\m_{n}(\d\gu_*^-).
\]
The estimate \er{aspamun} implies
$$
\Big|d_\gu \m_{n}(\d\gu)-
{2\pi n\/3}\big(\sqrt 3\d p_{cn}+\d p_{sn}\big)
+{1\/\sqrt 3}\big(\sqrt 3\d q_{cn}+\d q_{sn}\big)\Big|
\le{C\|\gu\|_1\/n} .
$$
By substituting $\gu_*^-$ for $\gu$,
we obtain
$$
\Big|d_{\gu_*^-}\m_{n}(\d\gu_*^-)-
{2\pi n\/3}\big(\sqrt 3\d p_{cn}-\d p_{sn}\big)
-{1\/\sqrt 3}\big(\sqrt 3\d q_{cn}-\d q_{sn}\big)\Big|
\le{C\|\gu\|_1\/n}.
$$
The identity \er{dm-ndm*n} gives
$$
\Big|d_\gu \m_{-n}(\d\gu)+{2\pi n\/3}\big(\sqrt 3\d p_{cn}-\d p_{sn}\big)
+{1\/\sqrt 3}\big(\sqrt 3\d q_{cn}-\d q_{sn}\big)\Big|
\le {C\|\gu\|_1\/n},
$$
which yields \er{aspamun} for $n<0$.~\BBox

\section{\lb{Sec7} Gradients of the norming constants}
\setcounter{equation}{0}

\subsection{Norming constants}
We proved in \cite[Lm~4.4]{BK25} that
if $(\l,\gu)\in\mD\ts\cB(\ve)$,
then there exists the simple multiplier $\t_3$
satisfying the estimate \er{lom}.
The function $\t_3(\l,\gu)$ is analytic on the domain
$\mD\ts\cB_\C(\ve)$, real for real $(\l,\gu)$, and positive
for all $\l>1$. Moreover, it satisfies
\[
\lb{idpataupau}
{\pa\t_3(\l)\/\pa\gu}
=-{{\pa D(\l,\t)\/\pa\gu}\/{\pa D(\l,\t)\/\pa \t}}\bigg|_{\t=\t_3(\l)},\qq
{\pa D(\l,\t)\/\pa\gu}={\pa T(\l)\/\pa\gu}\t^2-{\pa \wt T(\l)\/\pa\gu}\t,
\]
\[
\lb{idpataupau1}
{\pa D\/\pa \t}\Big|_{\t=\t_3}
=-(\t_1-\t_3)(\t_2-\t_3)
=-3\t_3^2+2T\t_3-\wt T,
\]
where we used the identity \er{cM}.

Let $n\in\N,\gu\in\cB(\ve)$.
The definition \er{defnf} and the identities \er{ef'1} imply
\[
\lb{idncr}
h_{sn}
=8(\pi n)^2\log\Big((-1)^n{\wt\varphi_n'(1)\/\wt\varphi_n'(0)}\t_3^{-{1\/2}}(\wt\m_n)\Big)
=8(\pi n)^2\log\Big((-1)^{n+1}{ y_3(1,\wt\m_n)\/\wt y_3(1,\wt\m_n)}\t_3^{-{1\/2}}(\wt\m_n)\Big).
\]
where $\wt\varphi_n(x)=\wt\varphi(x,\wt\m_n)$
is the eigenfunction of the operator $\wt\cL$
corresponding to the eigenvalue $\wt\m_n$ and given by \er{idef3p},
where $ y_j$ are replaced by $\wt y_j$.
This identity gives
\[
\lb{idgradhsn}
{\pa h_{sn}\/\pa\gu(t)}=
8(\pi n)^2\Big(\mA_n{\pa\wt\m_n\/\pa\gu(t)}
+\mB_n(t)\Big),
\qq t\in[0,1],
\]
where
\[
\lb{dfegA}
\mA_n={\dot y_3(1,\wt\m_n)\/ y_3(1,\wt\m_n)}
-{\dot{\wt y_3}(1,\wt\m_n)\/\wt y_3(1,\wt\m_n)}-{\dot\t_3(\wt\m_n)\/2\t_3(\wt\m_n)},
\]
\[
\lb{dfegB}
\mB_n(t)
={1\/ y_3(1,\wt\m_n)}{\pa  y_3(1,\wt\m_n)\/\pa\gu(t)}
-{1\/\wt y_3(1,\wt\m_n)}{\pa \wt y_3(1,\wt\m_n)\/\pa\gu(t)}
-{1\/2\t_3(\wt\m_n)}{\pa \t_3(\wt\m_n)\/\pa\gu(t)}.
\]

\subsection{The unperturbed case}

In the unperturbed case $\gu=0$ we have $\t_1^0(\l)=e^{\o z}$,
$\t_2^0(\l)=e^{\o^2 z}$, and
$\t_3^0(\l)=e^z$, then the identities \er{fremn} give
\[
\lb{tau30}
\t_1^0(\iota_n)=\t_2^0(\iota_n)=(-1)^ne^{-\nu n},
\qq
\t_3^0(\iota_n)=e^{2\nu n},
\qq
\dot\t_3^0(\iota_n)={e^{2\nu n}\/(2\pi n)^2},
\]
for all $n\in\N$.
The identities \er{derTrT=0} and \er{idpataupau} imply
\[
\lb{grmult}
{\pa\t_3(\l)\/\pa\gu(t)}\Big|_{\gu=0}=0,
\]
for all $(t,\l)\in[0,2]\ts\mD$.
In the following lemma
we calculate the gradient of the function $h_{sn}(\gu)$ at $\gu=0$.

\begin{lemma} Let $n\in\N$ and let  $t\in[0,1]$.
Then the function $y_3$ satisfies
\[
\lb{grvp3p=0}
{\pa  y_3(1,\iota_n)\/\pa\gu(t)}\Big|_{\gu=0}
=-{e^{2\nu n}\/4\sqrt 3(\pi n)^3}\Big(
\a_n(t)-(-1)^nc_{2n}(t)e^{-3\nu n}
;
{\sqrt 3\/4\pi n}
\big(\b_n(t)+2(-1)^nc_{2n}(t)e^{-3\nu n}\big)\Big),
\]
\[
\lb{grwtvp3p=0}
{\pa\wt y_3(1,\iota_n)\/\pa\gu(t)}\Big|_{\gu=0}
={(-1)^ne^{\nu n}\/4\sqrt 3(\pi n)^3}\Big(
\a_n(t)-c_{2n}(t);
{\sqrt 3\/4\pi n}\big(\b_n(t)+2c_{2n}(t)\big)\Big),
\]
where $c_n$ is given by \er{defcnsn},
\[
\lb{alnben}
\a_n(t)=-c_n(t)\big(e^{-3\nu nt}+(-1)^ne^{-3\nu n(1-t)}\big),
\qq
\b_n(t)=a_{-n}(t)e^{-3\nu nt}+(-1)^na_n(t)e^{-3\nu n(1-t)},
\]
 $a_n(t)$ is given by \er{anbn}.
The functions $\mA_n^0=\mA_n|_{\gu=0}$ and $\mB_n^0=\mB_n|_{\gu=0}$
satisfy
\[
\lb{idmA0}
\mA_n^0={ 3\/8(\pi n)^2},\qq
\mB_n^0(t)
=-{\sqrt3c_{2n}(t)\gv_n\/2(\pi n)^2},
\]
where $\gv_n$ is given by \er{defgvn}.
Moreover, the functions $h_{sn}$ satisfy
\[
\lb{grhsnp=0}
{\pa h_{sn}\/\pa\gu(t)}\Big|_{\gu=0}
=\sqrt 3a_{-2n}(t)\gv_n.
\]

\end{lemma}

\no {\bf Proof.}
Let $n\in\N$ and let $t\in[0,1]$. The identities \er{dotvp30mn} give
$$
{\dot y_{3,n}^0(1)\/ y_{3,n}^0(1)}
={ 1\/(2\pi n)^2}\Big(1-{\sqrt3\/\pi n}\Big),
\qqq
{\dot{\wt y_{3,n}^0}(1)\/\wt y_{3,n}^0(1)}
=-{1\/(2\pi n)^2}\Big(1+{\sqrt3\/\pi n}\Big).
$$
Substituting these identities and the identity \er{tau30} into the definition
\er{dfegA}, we obtain the first identity in
\er{idmA0}.

We prove \er{grvp3p=0}.
The identities
\er{derlfm} and \er{derpsifm} and  the definitions \er{mtrH} imply
\[
\lb{idgrvp3unp}
{\pa  y_3(1,\iota_n)\/\pa\gu(t)}\Big|_{\gu=0}
=-\big(\F_0(1-t,\iota_n)( J_p, J_q)\F_0(t,\iota_n)\big)_{13}
=-\big(A_n(t);B_n(t)\big),
\]
where $\F_0$ is given by \er{idX0cU},
$$
\begin{aligned}
A_n(t)=\F_{0,12}(1-t,\iota_n)\F_{0,13}(t,\iota_n)
+\F_{0,13}(1-t,\iota_n)\F_{0,23}(t,\iota_n)
\\
=y_{2,n}^0(1-t) y_{3,n}^0(t)+ y_{3,n}^0(1-t) y_{2,n}^0(t),
\end{aligned}
$$
$$
B_n(t)=\F_{0,13}(1-t,\iota_n)\F_{0,13}(t,\iota_n)
=y_{3,n}^0(1-t) y_{3,n}^0(t) .
$$
The identities \er{vp23tunp},  \er{vp23tunp1},
and $a_n(1-t)=(-1)^na_{-n}(t)$ give
\[
\lb{An(t)}
\begin{aligned}
A_n(t)={e^{2\nu n}\/\sqrt 3(2\pi n)^3}
\Big(\big(1+(-1)^na_{-n}(t)e^{-3\nu n(1-t)}\big)
\big(1+a_{-n}(t)e^{-3\nu nt}\big)
\\
+\big(1+(-1)^na_n(t)e^{-3\nu n(1-t)}\big)
\big(1+a_n(t)e^{-3\nu nt}\big)\Big)=
{e^{2\nu n}\/4\sqrt 3(\pi n)^3}(\a_n(t)-(-1)^nc_{2n}(t)e^{-3\nu n})
+\const,
\end{aligned}
\]
\[
\lb{Bn(t)}
\begin{aligned}
B_n(t)=
{e^{2\nu n}\/(2\pi n)^4}\big(1+(-1)^na_n(t)e^{-3\nu n(1-t)}\big)
\big(1+a_{-n}(t)e^{-3\nu nt}\big)
\\
={e^{2\nu n}\/(2\pi n)^4}(\b_n(t)+2(-1)^nc_{2n}(t)e^{-3\nu n})
+\const,
\end{aligned}
\]
where $\const$ is a value that does not depend on the variable $t$,
 and we used the identities
$$
a_n(t)+a_{-n}(t)=-2c_n(t),
\qq a_n^2(t)+a_{-n}^2(t)=4-2c_{2n}(t),\qq
a_n(t)a_{-n}(t)=2c_{2n}(t)-1.
$$
Substituting the identities \er{An(t)} and \er{Bn(t)}
into \er{idgrvp3unp}, we obtain
\er{grvp3p=0}.

The identity \er{grwtX(1)} implies
\[
\lb{pawty3pagu}
{\pa\wt y_3(1,\iota_n)\/\pa\gu(t)}\Big|_{\gu=0}=
{\pa\wt \F_{13}(1,\iota_n)\/\pa\gu(t)}\Big|_{\gu=0}
=-\big(\wt \F_0(1-t,\iota_n)( J_p,- J_q)\wt \F_0(t,\iota_n)\big)_{13}
=-\big(\wt A_n(t);\wt B_n(t)\big),
\]
where
$$
\begin{aligned}
\wt A_n(t)=\wt \F_{0,12}(1-t,\iota_n)\wt \F_{0,13}(t,\iota_n)
+\wt \F_{0,13}(1-t,\iota_n)\wt \F_{0,23}(t,\iota_n)
\\
=\wt y_{2,n}^0(1-t) \wt y_{3,n}^0(t)+
\wt y_{3,n}^0(1-t)\wt  y_{2,n}^0(t)=A_{-n}(t),
\end{aligned}
$$
$$
\wt B_n(t)=-\wt \F_{0,13}(1-t,\iota_n)\wt \F_{0,13}(t,\iota_n)
=-\wt y_{3,n}^0(1-t) \wt y_{3,n}^0(t)=-B_{-n}(t),
$$
here we used the identities \er{symyn0}.
The identities \er{An(t)}, \er{Bn(t)}, and
$$
\a_{-n}(t)=(-1)^ne^{3\nu n}\a_{n}(t),\qq
\b_{-n}(t)=(-1)^ne^{3\nu n}\b_{n}(t),
$$
give
$$
\wt A_n(t)=
{(-1)^{n+1}e^{\nu n}(\a_{n}(t)-c_{2n}(t))\/4\sqrt 3(\pi n)^3}
+\const,
\qq
\wt B_n(t)=
{(-1)^{n+1}e^{\nu n}(\b_{n}(t)+2c_{2n}(t))\/(2\pi n)^4}
+\const.
$$
Substituting these identities into \er{pawty3pagu}
we obtain \er{grwtvp3p=0}.
Substituting the identities \er{vp23tunp1}, \er{grmult}, and \er{grwtvp3p=0}
into the definition \er{dfegB}, we obtain the second identity in \er{idmA0}.

Substituting the identities \er{idmA0}
 and \er{grmuunp} into \er{idgradhsn}
we obtain
$$
{\pa h_{sn}\/\pa\gu(t)}\Big|_{\gu=0}
=\big(3b_{-2n}(t)-4\sqrt3c_{2n}(t)\big)\gv_n,
$$
where $\gv_n$ and $b_{n}(t)$ are given by \er{defgvn} and \er{bn}.
The identity $\sqrt3b_{-2n}(t)-4c_{2n}(t)=a_{-2n}(t)$
 yields \er{grhsnp=0}.~\BBox

 \subsection{Asymptotics of the multiplier}
We determine the following asymptotics for the multiplier $\t_3$.

\begin{lemma}
\lb{LmGD}
If $\gu\in\cB(\ve)$,
then
\[
\lb{esttau123}
 C_1e^{\Re z}\le|\t_3(\l)|\le C_2e^{\Re z},\qq
\sup_{t\in[0,1]}\Big(\Big|{\pa\t_3(\l)\/\pa p(t)}\Big|
+|z|\Big|{\pa\t_3(\l)\/\pa q(t)}\Big|\Big)\le {C\|\gu\|_1\/|z|^{3}}e^{\Re z},
\]
for all $\l\in\mD$ and
for some positive $C,C_1,C_2$. Moreover,
\[
\lb{asderltau3}
\big|\t_3(\wt\m_n)-e^{2\nu n}\big|
\le {C\|\gu\|_1\/n^2}e^{2\nu n},
\qqq
\Big|\dot\t_3(\wt\m_n)
-{e^{2\nu n}\/(2\pi n)^2}\Big|\le{C\|\gu\|_1\/n^4}e^{2\nu n},
\]
\[
\lb{dertau3-0}
\begin{aligned}
{1\/|\t_3(\wt\m_n)|}\sup_{t\in[0,1]}\Big(\Big|{\pa\t_3(\wt\m_n)\/\pa p(t)}\Big|+
n\Big|{\pa\t_3(\wt\m_n)\/\pa q(t)}\Big|\Big)
\le{C\|\gu\|_1\/n^{3}},
\end{aligned}
\]
for all $n\in\N$ and for some  $C>0$.
\end{lemma}

\no {\bf Proof.}
Let $\gu\in\cB(\ve)$. The estimate \er{lom} yields the first estimates
in \er{esttau123}.
Gershgorin's theorem (see, e.g., \cite[Ch~4.3]{BK25}) gives
$$
|\t_1(\l)|\le C |e^{\o z}|\le Ce^{{-\Re z+\sqrt3|\Im z|\/2}},\qq
|\t_2(\l)|\le C |e^{\o^2 z}|\le Ce^{{-\Re z+\sqrt3|\Im z|\/2}},
$$
for some $C>0$ and
for all $\l\in\C\sm\cup_{n\in\Z}\cD_{n}$. The function $\t_1+\t_2$ is
analytic on $\mD$, then, due to Cauchy's formula,
$$
|\t_1(\l)+\t_2(\l)|\le Ce^{{-\Re z+\sqrt3|\Im z|\/2}},
$$
for some $C>0$ and for all $\l\in\mD$.
The identity \er{idpataupau1} and the first estimates in \er{esttau123}
imply
\[
\lb{estderDb}
\Big|{\pa D(\l,\tau)\/\pa \t}\big|_{\t=\t_3(\l)}\Big|
=|\t_1(\l)-\t_3(\l)||\t_2(\l)-\t_3(\l)|
=\Big|{1\/\t_3(\l)}-\big(\t_1(\l)+\t_2(\l)\big)\t_3(\l)+\t_3^2(\l)\Big|\ge Ce^{2\Re z },
\]
for some $C>0$ and for all $\l\in\mD$.
The identity \er{idpataupau}, the estimates \er{estgrT}
and the first estimate in \er{esttau123} imply
\[
\lb{estchisl}
\begin{aligned}
\sup_{t\in[0,1]}\Big(\Big|{\pa D(\l,\tau)\/\pa p(t)}\big|_{\tau=\t_3(\l)}\Big|
+|z|\Big|{\pa D(\l,\tau)\/\pa q(t)}\big|_{\tau=\t_3(\l)}\Big|\Big)
\\
=\sup_{t\in[0,1]}\Big(\Big|{\pa T(\l)\/\pa p(t)}\t_3^2(\l)
-{\pa \wt T(\l)\/\pa p(t)}\t_3(\l)\Big|+
|z|\Big|{\pa T(\l)\/\pa q(t)}\t_3^2(\l)
-{\pa \wt T(\l)\/\pa q(t)}\t_3(\l)\Big|\Big)
\le e^{3\Re z }{C\|\gu\|_1\/|z|^{3}},
\end{aligned}
\]
for some $C>0$ and for all $\l\in\mD$.
Substituting the estimates \er{estchisl} and \er{estderDb}
into the identity \er{idpataupau}, we obtain
 the second estimate in \er{esttau123}. The estimates  \er{esttau123}
imply \er{dertau3-0}.

We prove  \er{asderltau3}.
Let $n\in\N$. Then for $f=\t_3$ or $f=\dot\t_3$ we have
\[
\lb{estt3-t30pr}
|f(\wt\m_n)-f_0(\iota_n)|\le
|f(\wt\m_n)-f_0(\wt\m_n)|
+|f_0(\wt\m_n)-f_0(\iota_n)|,
\]
where $f_0=\t_3^0$ or $f_0=\dot\t_3^0$, respectively. Cauchy's formula gives
$$
\t_3(\wt\m_n)-\t_3^0(\wt\m_n)
={1\/2\pi i }\oint_{|\z-{2\nu n}|=1} {\t_3(\z^3)-\t_3^0(\z^3)\/\z-z_n}d\z,
\qqq z_n=\wt\m_n^{1\/3}>0,
$$
$$
\dot\t_3(\wt\m_n)-\dot\t_3^0(\wt\m_n)
={1\/3z^2}{d(\t_3(\l)-\t_3^0(\l))\/dz}\Big|_{\l=\wt\m_n}
={1\/6\pi i z_n^2}\oint_{|\z-{2\nu n}|=1}
{\t_3(\z^3)-\t_3^0(\z^3)\/(\z-z_n)^2}d\z.
$$
These identities and  the estimate \er{lom} imply
\[
\lb{estdotmtpr}
|\t_3(\wt\m_n)-\t_3^0(\wt\m_n)|
\le C\max_{\l\in\pa\cD_n}|\t_3(\l)-\t_3^0(\l)|
\le{C\|\gu\|_1^2\/n^4}e^{2\nu n},\qq
|\dot\t_3(\wt\m_n)-\dot\t_3^0(\wt\m_n)|
\le{C\|\gu\|_1^2\/n^6}e^{2\nu n},
\]
for some $C>0$.
Moreover, the estimate \er{locmn} yields
$$
\big|\t_3^0(\wt\m_n)-\t_3^0(\iota_n)\big|
=|e^{z_n}-e^{2\nu n}|
\le C|z_n-2\nu n|e^{2\nu n}
\le {C\|\gu\|_1\/n^2}e^{2\nu n},
$$
$$
\big|\dot\t_3^0(\wt\m_n)-\dot\t_3^0(\iota_n)\big|
={1\/3}\Big|{e^{z_n}\/z_n^2}-{e^{2\nu n}\/(2\nu n)^2}\Big|
\le {1\/3}\max_{z\in[2\nu n,z_n]}\Big|{d\/dz}{e^z\/z^2}\Big|
|z_n-2\nu n|
\le {C\|\gu\|_1\/n^4}e^{2\nu n},
$$
for some $C>0$.
Substituting these estimates and the estimates \er{estdotmtpr}
into  \er{estt3-t30pr} and using \er{tau30}, we obtain \er{asderltau3}.~\BBox

\section{\lb{Sec8} Asymptotics of the gradients of the norming constants}
\setcounter{equation}{0}

\subsection{Preliminaries}
The definitions \er{mtrH}, \er{4g.Om}, and \er{defmaZ}
give
\[
\lb{JpJqUZ}
JJ_pJ=-J_p^\top,\ \ JJ_qJ=J_q^\top,\ \
 Z^{-1} J_p Z={ J_p\/z},\ \
 Z^{-1} J_q Z={ J_q\/z^2},\ \
U^{-1} J_pU={\O_p\/3},\ \
U^{-1} J_qU={\O_q\/3},
\]
where
\[
\lb{Ompq}
\O_p=\ma 2\o^2&-\o&-1\\-\o^2&2\o&-1\\-\o^2&-\o&2\am,\qq
\O_q=\ma \o&\o&\o\\\o^2&\o^2&\o^2\\1&1&1\am.
\]

Let $x\in[0,1],n\in\N$.
Introduce the functions
\[
\lb{cWcW00}
 P_n^1(x)=W_1(1-x,x,\iota_n)\O_pW_1(x,\iota_n),\qq
 Q_n^1(x)=W_1(1-x,x,\iota_n) \O_q W_1(x,\iota_n),
\]
\[
\lb{wtcWn0}
\wt P_n^1(x)=\wt W_1(1-x,x,\iota_n)\O_p^\top\wt W_1(x,\iota_n),
\qq
\wt Q_n^1(x)=\wt W_1(1-x,x,\iota_n)\O_q^\top\wt W_1(x,\iota_n),
\]
where
$$
W_1(x,t,\l,\gu)=W_1(x,\l,\gu(\cdot+t)),\qq
\wt W_1(x,t,\l,\gu)=\wt W_1(x,\l,\gu(\cdot+t)),
$$
$W_1,\wt W_1$ are given by  \er{W1wtW1} and \er{defwtW0}.

\begin{lemma} Let $\gu\in\cB(\ve)$, let $j,k=1,2,3$, and let $n\in\N$. Then
\[
\lb{estWn1-Wn00}
e^{-2\nu n}\sup_{x\in[0,1]}\big| Q_{n,jk}^1(x)-\o^je_{n,jk}(x)\big|+
e^{-\nu n}\sup_{x\in[0,1]}\big|\wt Q_{n,jk}^1(x)-\o^k e^{-1}_{n,jk}(x)\big|
\le {C\|\gu\|\/n},
\]
\[
\lb{estintcW-cW0}
e^{-2\nu n}\Big|\int_0^1 \big( Q_{n,jk}^1(x)-\o^je_{n,jk}(x)\big)dx\Big|+
e^{-\nu n}\Big|\int_0^1 \big(\wt Q_{n,jk}^1(x)-\o^k  e^{-1}_{n,jk}(x)\big)dx
\le {C\|\gu\|_1\/n^2},
\]
\[
\lb{estintP1-e}
e^{-2\nu n}\Big|\int_0^1 \big(P_{n,jk}^1(x)+\O_{p,jk}e_{n,jk}(x)\big)\d p(x)dx\Big|+
e^{-\nu n}\Big|\int_0^1\Big(\wt P_{n,jk}^1(x)+{\O_{p,kj}\/e_{n,jk}(x)}\Big)\d p(x)dx\Big|
\le {C\|\gu\|_1\/n^2},
\]
for all $\d p\in\cH_1$ and for some $C>0$, where
\[
\lb{cWn0wtcWn0}
e_{n,jk}(x)=e^{2n\nu(1-x)\o^j+2n\nu x\o^k}.
\]

\end{lemma}

\no {\bf Proof.} Let $j,k=1,2,3$.
The definitions \er{W1wtW1}, \er{defwtW0}, \er{Ompq}, \er{cWcW00},
and \er{wtcWn0} give
$$
\begin{aligned}
P_{n,jk}^1(x)
=-\o^{2k}\cE_{n,jk}(x),\qq
\wt P_{n,jk}^1(x)
=-\o^{2j} \cE^{-1}_{n,jk}(x),\qq j\ne k,
\\
P_{n,jj}^1=2\o^{2j}e_{n,jj}=2\o^{2j}e^{2n\nu\o^j},\qq
\wt P_{n,jj}^1=2\o^{2j} e^{-1}_{n,jj}=2\o^{2j}e^{-2n\nu\o^j},
\end{aligned}
$$
$$
 Q_{n,jk}^1(x)
=\o^j\cE_{n,jk}(x),\qq
\wt Q_{n,jk}^1(x)
=\o^k \cE^{-1}_{n,jk}(x),
$$
where $e_{n,jk}$ are given by \er{cWn0wtcWn0},
$$
\cE_{n,jk}(x)=e_{n,jk}(x)e^{-p_{n,jk}(x)},\qq
p_{n,jk}(x)={1\/6\nu n}\Big(\o^{2j}\int_0^{1-x}p(s+x)ds
+\o^{2k}\int_0^xp(s)ds\Big),\qq p_{n,jj}=0.
$$
Note that $P_{n,jj}^1$, $\wt P_{n,jj}^1$, and $e_{n,jj}$ are constants.
This yields \er{estintP1-e} for $j=k$. Moreover,
$ Q_{n,jj}^1=\o^je_{n,jj}$ and
$\wt Q_{n,jj}^1=\o^j e^{-1}_{n,jj}$. Therefore, \er{estWn1-Wn00}
and \er{estintcW-cW0} for $j=k$ are fulfilled. Consider the case $j\ne k$.
The estimate \er{estexpomx}
gives
\[
\lb{estP1-cG}
e^{-2\nu n}| P_{n,jk}^1(x)+\o^{2k}f_{n,jk}(x)|
+e^{-\nu n}|\wt P_{n,jk}^1(x)dx+\o^{2j}\wt f_{n,jk}(x)|
\le {C\|\gu\|^2\/n^2},
\]
\[
\lb{estWn1-Wn0}
 e^{-2\nu n}| Q_{n,jk}^1(x)-\o^jf_{n,jk}(x)|+
e^{-\nu n}|\wt Q_{n,jk}^1(x)-\o^k\wt f_{n,jk}(x)|\le {C\|\gu\|^2\/n^2},
\]
for some $C>0$, where
\[
\lb{defcGnjk}
f_{n,jk}(x)=e_{n,jk}(x)\big(1-p_{n,jk}(x)\big),
\qq
\wt f_{n,jk}(x)= e^{-1}_{n,jk}(x)\big(1+p_{n,jk}(x)\big).
\]
The estimates \er{estWn1-Wn0}
and the definitions \er{defcGnjk} yield \er{estWn1-Wn00}. Integrating by parts, we obtain
\[
\lb{estcG11}
e^{-2\nu n}\Big|\int_0^1\big(f_{n,jk}(x)- e_{n,jk}(x)\big)dx\Big|
+e^{-\nu n}\Big|\int_0^1\big(\wt f_{n,jk}(x)- e^{-1}_{n,jk}(x)\big)dx\Big|
\le {C\|\gu\|_1\/n^2},
\]
\[
\lb{estintcG-e}
e^{-2\nu n}\Big|\int_0^1\big(f_{n,jk}(x)-e_{n,jk}(x)\big)
\d p(x)dx\Big|+
e^{-\nu n}\Big|\int_0^1\big(\wt f_{n,jk}(x)- e^{-1}_{n,jk}(x)\big)
\d p(x)dx\Big|\le {C\|\gu\|_1\/n^2},
\]
for some $C>0$, where $\d p\in\cH_1$.
The estimates \er{estcG11} and \er{estWn1-Wn0} imply \er{estintcW-cW0}.
The estimates \er{estP1-cG} and \er{estintcG-e} yield \er{estintP1-e}.~\BBox

\medskip

Let $n\in\N,x\in[0,1]$.
Introduce the functions
\[
\lb{cWcW0}
 P_n(x)=W(1-x,x,\wt\m_n)\O_pW(x,\wt\m_n),\qq
 Q_n(x)=W(1-x,x,\wt\m_n) \O_q W(x,\wt\m_n),
\]
\[
\lb{wtcWn}
\wt P_n(x)=\wt W(1-x,x,\wt\m_n)\O_p^\top \wt W(x,\wt\m_n),\qq
\wt Q_n(x)=\wt W(1-x,x,\wt\m_n)\O_q^\top\wt W(x,\wt\m_n),
\]
where
$$
W(x,t,\l,\gu)=W(x,\l,\gu(\cdot+t)),\qq
\wt W(x,t,\l,\gu)=\wt W(x,\l,\gu(\cdot+t)),
$$
$W,\wt W$ are given by \er{W1wtW1} and \er{wtW1wtW1}.
Below we need the following estimates.

\begin{lemma}
Let $\gu\in\cB(\ve)$, let $j,k=1,2,3$, and let $n\in\N$. Then
\[
\lb{estCW-Wn00}
e^{-2\nu n}\sup_{x\in[0,1]}\big| Q_{n,jk}(x)-\o^je_{n,jk}(x)\big|+
e^{-\nu n}\sup_{x\in[0,1]}\big|\wt Q_{n,jk}(x)-\o^k e^{-1}_{n,jk}(x)\big|\le
{C\|\gu\|_1\/n},\qq
\]
\[
\lb{estcW112}
e^{-2\nu n}\Big|\int_0^1 \big( Q_{n,jk}(x)-\o^je_{n,jk}(x)\big)dx\Big|
+e^{-\nu n}\Big|\int_0^1 \big(\wt Q_{n,jk}(x)-\o^k e^{-1}_{n,jk}(x)\big)dx\Big|
\le {C\|\gu\|_1\/n^2},
\]
\[
\lb{estcVOe}
e^{-2\nu n}\Big|\int_0^1\big( P_{n,jk}(x)+\O_{p,jk}e_{n,jk}(x)\big)\d p(x)dx\Big|
+e^{-\nu n}\Big|\int_0^1\Big(\wt P_{n,jk}(x)+{\O_{p,kj}\/e_{n,jk}(x)}\Big)\d p(x)dx\Big|
\le {C\|\gu\|_1\/n^2},
\]
for all $\d p\in\cH_1$ and for some $C>0$,
where $e_{n,jk}$ are given by \er{cWn0wtcWn0}.
\end{lemma}

\no {\bf Proof.} Let $x\in[0,1],n\in\N$.
Let  $f= P_n,f_0= P_n^1,g=\O_p$ or $f= Q_n,f_0= Q_n^1,g=\O_q$.
Then the definitions \er{cWcW00} and \er{cWcW0} give
$$
\begin{aligned}
\big|f(x)-f_0(x)\big|
\le
\big|W(1-x,x,\wt\m_n)-W_1(1-x,x,\iota_n)\big|\big|gW(x,\wt\m_n)\big|
\\
+\big|W_1(1-x,x,\iota_n) g\big|\big|W(x,\wt\m_n)-W_1(x,\iota_n)\big|.
\end{aligned}
$$
Let  $\wt f=\wt P_n,\wt f_0= \wt P_n^1,\wt g=\O_p^\top$ or
$\wt f= \wt Q_n,\wt f_0=\wt Q_n^1,\wt g=\O_q^\top$.
Then the definitions \er{wtcWn0} and \er{wtcWn} give
$$
\begin{aligned}
\big|\wt f(x)-\wt f_0(x)\big|
\le
\big|\wt W(1-x,x,\wt\m_n)-\wt W_1(1-x,x,\iota_n)\big|\big|\wt g\wt W(x,\wt\m_n)\big|
\\
+\big|\wt W_1(1-x,x,\iota_n) \wt g\big|\big|\wt W(x,\wt\m_n)-\wt W_1(x,\iota_n)\big|.
\end{aligned}
$$
The estimates \er{estX} and \er{asX-X0atmu0}  yield
\[
\lb{estCW-Wn0}
\big|f(x)-f_0(x)\big|\le e^{2\nu n}{C\|\gu\|_1\/n^{2}},\qq
\big|\wt f(x)-\wt f_0(x)\big|\le e^{\nu n}{C\|\gu\|_1\/n^{2}}.
\]
The estimates \er{estWn1-Wn00}, \er{estintcW-cW0}, \er{estintP1-e}, and \er{estCW-Wn0}
imply \er{estCW-Wn00}, \er{estcW112}, and  \er{estcVOe}.~\BBox

\subsection{Asymptotics of the solution $ y_3$.}
We prove the following estimates for the fundamental solution $ y_3$.

\begin{lemma}
Let $\gu\in\cB(\ve)$ and let $n\in\N$.
Then the function $ y_3$ satisfies
\[
\lb{estvp3-vpa30}
\big| y_3(1,\wt\m_n)- y_{3,n}^0(1)\big|
\le e^{{2\nu n} }{C\|\gu\|_1\/n^{4}},\qq
\big|\wt y_3(1,\wt\m_n)-\wt y_{3,n}^0(1)\big|
\le e^{{\nu n} }{C\|\gu\|_1\/n^{4}},
\]
\[
\lb{asdotvp3}
\begin{aligned}
\big|\dot y_3(1,\wt\m_n)-
\dot y_{3,n}^0(1)\big|
\le e^{2\nu n}{C\|\gu\|_1\/n^6},\qq
\big|\dot{\wt y_3}(1,\wt\m_n)-
\dot{\wt y_{3,n}^0}(1)\big|
\le e^{\nu n}{C\|\gu\|_1\/n^6},
\end{aligned}
\]
\[
\lb{esfdervpvp0}
\sup_{x\in[0,1]}\Big|{\pa  y_3(1,\wt\m_n)\/\pa \gu(x)}
-{\pa  y_3(1,\iota_n)\/\pa \gu(x)}\big|_{\gu=0}\Big|\le
e^{2\nu n}{C\|\gu\|_1\/n^{5}},
\]
\[
\lb{esfderwtvpvp0}
\sup_{x\in[0,1]}\Big|{\pa \wt y_3(1,\wt\m_n)\/\pa \gu(x)}
-{\pa \wt y_3(1,\iota_n)\/\pa \gu(x)}\big|_{\gu=0}\Big|\le
e^{\nu n}{C\|\gu\|_1\/n^{5}},
\]
for some $C>0$.
\end{lemma}

\no {\bf Proof.}
The estimate \er{estXjkmu-mu0} and the identities $y_3=\F_{13},\wt y_3=\wt\F_{13}$
yield \er{estvp3-vpa30}.
We prove \er{asdotvp3}. Let $n\in\N$.
 The identities  \er{derlfm}, \er{idXcX},
and \er{JpJqUZ} imply
\[
\lb{idZ-1dotFZ}
\dot \F(1,\wt\m_n)
=\int_0^1  \F(1-x,x,\wt\m_n) J_q\F(x,\wt\m_n)dx
={1\/3z_n^2}\int_0^1 Z(\wt\m_n) U Q_n(x)U^{-1}Z^{-1}(\wt\m_n)dx,
\]
where $z_n=\wt\m_n^{1\/3}, Q_n$ are given by \er{cWcW0}.
The identity \er{idZ-1dotFZ} and the definitions \er{defmaZ} imply
\[
\lb{iddoty3}
\dot y_3(1,\wt\m_n)=\dot\F_{13}(1,\wt\m_n)
={1\/9z_n^4} \sum_{j,k=1}^3\o^k\int_0^1 Q_{n,jk}(x)dx,
\]
\[
\lb{iddoty30}
\dot y_{3,n}^0(1)=\dot y_3(1,\wt\m_n)|_{\gu=0}=
{1\/(2\pi n)^4} \sum_{j,k=1}^3\o^k\o^j
\int_0^1 e_{n,jk}(x)dx.
\]
Moreover, the identities \er{symXM}, \er{derlfm}, and \er{repwtX} yield
\[
\lb{dotwtfi}
\begin{aligned}
\dot{\wt \F}(1,\l)=\int_0^1 {\wt \F}(1-t,t,\l) J_q{\wt \F}(t,\l)dt
\\
=\int_0^1 J Z^{-1}(\l) (U^{-1})^\top\wt W(1-t,t,\l)U^\top Z(\l)
J J_qJ Z^{-1}(\l)
(U^{-1})^\top\wt W(t,\l)U^\top Z(\l) Jdt,
\end{aligned}
\]
for all $\l\in\C$.
The identities \er{JpJqUZ} imply
\[
\lb{UtZJJqJZUt}
U^\top Z J J_qJ Z^{-1} (U^{-1})^\top
=U^\top Z J_q^\top Z^{-1} (U^{-1})^\top
={1 \/z^2}U^\top J_q^\top (U^{-1})^\top
={\O_q^\top\/3z^2}.
\]
The identity \er{dotwtfi} gives
$$
\dot{\wt \F}(1,\l)
={1\/3z^2}\int_0^1 J Z^{-1}(\l)
(U^{-1})^\top\wt W(1-t,t,\l)\O_q^\top\wt W(t,\l)U^\top Z(\l) Jdt,
$$
for all $\l\in\C$, which yields
$$
\dot{\wt \F}(1,\wt\m_n)
={1\/3z_n^2}\int_0^1 J Z^{-1}(\wt\m_n) (U^{-1})^\top
\wt Q_n(t)U^\top Z(\wt\m_n) Jdt,
$$
where $\wt Q_n$ is given by \er{wtcWn}.
Then the definitions \er{defmaZ} imply
\[
\lb{iddotwty3wtWn}
\dot{\wt y_3}(1,\wt\m_n)=\dot{\wt \F}_{13}(1,\wt\m_n)
={1\/9z_n^4}\sum_{j,k=1}^3\o^j\int_0^1\wt Q_{n,jk}(t)dt,
\]
\[
\lb{iddotwty30}
\dot{\wt y_{3,n}^0}(1)=\dot{\wt y_3}(1,\wt\m_n)|_{\gu=0}=
{1\/(2\pi n)^4} \sum_{j,k=1}^3\o^k\o^j
\int_0^1  e^{-1}_{n,jk}(x)dx.
\]
The identities \er{iddoty3}, \er{iddoty30}, \er{iddotwty3wtWn}, \er{iddotwty30},
 and  the estimates \er{locmn}, \er{estcW112} imply \er{asdotvp3}.

We prove \er{esfdervpvp0}. Let $n\in\N,x\in[0,1]$.
The identities \er{grX(1)}, \er{idXcX}, and  \er{JpJqUZ}
imply
$$
{\pa \F(1,\wt\m_n)\/\pa\gu(x)}
=-{1\/3z_n}Z(\wt\m_n) U\Big( P_n(x),{ Q_n(x)\/z_n}\Big)U^{-1}Z^{-1}(\wt\m_n),
\qq z_n=\wt\m_n^{1\/3},
$$
where $ P_n, Q_n$ are given by \er{cWcW0}.
This identity gives
\[
\lb{pay3pau}
{\pa  y_3(1,\wt\m_n)\/\pa\gu(x)}={\pa \F_{13}(1,\wt\m_n)\/\pa\gu(x)}
=-{1\/9z_n^3} \sum_{j,k=1}^3\o^k\Big( P_{n,jk}(x),{ Q_{n,jk}(x)\/z_n}\Big),
\]
\[
\lb{dery3p0}
{\pa  y_3(1,\iota_n)\/\pa \gu(x)}\Big|_{\gu=0}
=-{1\/9(2\nu n)^3}\sum_{j,k=1}^3\o^k\Big(\O_{p,jk},{\o^j\/2\nu n}\Big)
e_{n,jk}(x).
\]
The estimates \er{locmn} and \er{estCW-Wn00} give
\[
\lb{pavp3paq}
\bigg|{\pa  y_3(1,\wt\m_n)\/\pa q(x)}
-{\pa  y_3(1,\iota_n)\/\pa q(x)}\Big|_{\gu=0}\bigg|
\le e^{2\nu  n}{C\|\gu\|_1\/n^5},
\]
for some $C>0$.
If $\d p\in\cH_1$, then the estimate \er{estcVOe}
and the identities \er{pay3pau} and \er{dery3p0} imply
$$
\bigg|\int_0^1{\pa  y_3(1,\wt\m_n)\/\pa p(x)}\d p(x)-
\int_0^1{\pa  y_3(1,\iota_n)\/\pa p(x)}\Big|_{\gu=0}\d p(x)\bigg|
\le {e^{2\nu n}C\|\gu\|_1\/n^5}.
$$
This yields
$$
\bigg|{\pa  y_3(1,\wt\m_n)\/\pa p(x)}
-{\pa  y_3(1,\iota_n)\/\pa p(x)}\Big|_{\gu=0}\bigg|
\le e^{2\nu  n}{C\|\gu\|_1\/n^5},
$$
for some $C>0$. This estimate and the estimate \er{pavp3paq} give
\er{esfdervpvp0}.

We prove \er{esfderwtvpvp0}. Let $n\in\N,x\in[0,1]$.
The identities \er{repwtX} and \er{grwtX(1)} imply
\[
\lb{pawtfi1gu}
{\pa\wt \F(1,\wt\m_n)\/\pa\gu(x)}
=-J Z^{-1}(\l) (U^{-1})^\top\wt W(1-x,x,\l)
U^\top Z(\l) J (J_p,-J_q)J Z^{-1}(\l) (U^{-1})^\top
\wt W(x,\l)U^\top Z(\l) J.
\]
The identities \er{JpJqUZ} imply
$$
U^\top Z J J_pJ Z^{-1} (U^{-1})^\top=
-U^\top Z J_p^\top Z^{-1} (U^{-1})^\top=
-{1\/z}U^\top J_p^\top (U^{-1})^\top=
-{\O_p^\top\/3z}.
$$
Substituting this identity and the identity \er{UtZJJqJZUt}
into \er{pawtfi1gu}, we obtain
$$
{\pa\wt \F(1,\wt\m_n)\/\pa\gu(x)}
={1\/3z}J Z^{-1}(\l) (U^{-1})^\top\wt W(1-x,x,\l)
\Big(\O_p^\top,{\O_q^\top\/z}\Big)\wt W(x,\l)U^\top Z(\l) J.
$$
The definitions \er{wtcWn} imply
$$
{\pa\wt \F(1,\wt\m_n)\/\pa\gu(x)}
={1\/3z}J Z^{-1}(\l)(U^{-1})^\top
\Big(\wt P_n(x),{\wt Q_n(x)\/z}\Big)U^\top Z(\l) J.
$$
This identity gives
\[
\lb{pay3wtpau}
{\pa\wt  y_3(1,\wt\m_n)\/\pa\gu(x)}={\pa\wt \F_{13}(1,\wt\m_n)\/\pa\gu(x)}
={1\/9z_n^3} \sum_{j,k=1}^3\o^j\Big(\wt P_{n,jk}(x),{\wt Q_{n,jk}(x)\/z_n}\Big),
\]
\[
\lb{dery3wtp0}
{\pa \wt y_3(1,\iota_n)\/\pa \gu(x)}\Big|_{\gu=0}
={1\/9(2\nu n)^3} \sum_{j,k=1}^3\o^j\Big(\O_{p,kj},{\o^k\/2\nu n}\Big)
e^{-1}_{n,jk}(x).
\]
The estimates \er{locmn} and \er{estCW-Wn00} give
\[
\lb{pavp3wtpaq}
\bigg|{\pa  y_3(1,\wt\m_n)\/\pa q(x)}
-{\pa  y_3(1,\iota_n)\/\pa q(x)}\Big|_{\gu=0}\bigg|
\le e^{\nu  n}{C\|\gu\|_1\/n^5},
\]
for some $C>0$.
If $\d p\in\cH_1$, then the second estimate in \er{estcVOe}
and the identities \er{pay3wtpau} and \er{dery3wtp0} imply
$$
\bigg|\int_0^1{\pa \wt y_3(1,\wt\m_n)\/\pa p(x)}\d p(x)-
\int_0^1{\pa \wt y_3(1,\iota_n)\/\pa p(x)}\Big|_{\gu=0}\d p(x)\bigg|
\le {e^{\nu n}C\|\gu\|_1\/n^5}.
$$
This yields
$$
\bigg|{\pa \wt y_3(1,\wt\m_n)\/\pa p(x)}
-{\pa \wt y_3(1,\iota_n)\/\pa p(x)}\Big|_{\gu=0}\bigg|
\le e^{\nu  n}{C\|\gu\|_1\/n^5},
$$
for some $C>0$. This estimate and the estimate \er{pavp3wtpaq} give
\er{esfderwtvpvp0}.~\BBox

\subsection{Asymptotics of $\mA_n$ and $\mB_n$.}
We determine asymptotics of the sequences $\mA_n$ and $\mB_n$ given by
\er{dfegA} and \er{dfegB}.

\begin{lemma}
Let $\gu\in\cB(\ve)$ and let $n\in\N$.
Then

i) The functions $y_3,\wt y_3$, and $\tau_3$ satisfy
\[
\lb{asdotvpdevvp}
\Big|{\dot y_3(1,\wt\m_n)\/ y_3(1,\wt\m_n)}
-{\dot y_{3,n}^0(1)\/ y_{3,n}^0(1)}\Big|\le
{C\|\gu\|_1\/n^4},
\qqq
\Big|{\dot{\wt y_3}(1,\wt\m_n)\/\wt y_3(1,\wt\m_n)}
-{\dot{\wt y_{3,n}^0}(1)\/\wt y_{3,n}^0(1)}\Big|
\le{C\|\gu\|_1\/n^4},
\]
\[
\lb{dottau3-0}
\Big|{\dot\t_3(\wt\m_n)\/\t_3(\wt\m_n)}
-{\dot\t_3^0(\iota_n)\/\t_3^0(\iota_n)}\Big|\le{C\|\gu\|_1\/n^4},
\]
\[
\lb{asdervpdevvp}
\bigg|{1\/ y_3(1,\wt\m_n)}{\pa  y_3(1,\wt\m_n)\/\pa \gu(t)}
-{1\/ y_{3,n}^0(1)}{\pa  y_3(1,\iota_n)\/\pa \gu(t)}\Big|_{\gu=0}\bigg|\le
{C\|\gu\|_1\/n^3},
\]
\[
\lb{asdervpdevwtvp}
\bigg|{1\/\wt y_3(1,\wt\m_n)}{\pa \wt y_3(1,\wt\m_n)\/\pa \gu(t)}
-{1\/\wt y_{3,n}^0(1)}{\pa\wt  y_3(1,\iota_n)\/\pa \gu(t)}\Big|_{\gu=0}\bigg|\le
{C\|\gu\|_1\/n^3},
\]
for some $C>0$.

ii) The sequences $\mA_n$ and $\mB_n$,
given by  \er{dfegA} and \er{dfegB}, satisfy
\[
\lb{estL-Lo}
|\mA_n-\mA_n^0|\le {C\|\gu\|_1\/n^4},
\]
\[
\lb{estB-Bo}
\sup_{x\in[0,1]}|\mB_n(x)-\mB_n^0(x)|\le {C\|\gu\|_1\/n^3},
\]
for some $C>0$.
\end{lemma}

\no {\bf Proof.}
i) Let $n\in\N$. The identity \er{vp23tunp1}
and the estimate \er{estvp3-vpa30} imply
\[
\lb{estdoty30y3}
|y_{3,n}^0(1)|\ge {c_1e^{2\nu n}\/n^2},\qq
|y_3(1,\wt\m_n)|\ge |y_{3,n}^0(1)|-|y_3(1,\wt\m_n)-y_{3,n}^0(1)|
\ge {c_2e^{2\nu n}\/n^2},
\]
for some $c_1,c_2>0$.
Let $f_n=\dot y_3(1,\wt\m_n)$ and $f_n^0=\dot y_{3,n}^0(1)$,
or $f_n={\pa  y_3(1,\wt\m_n)\/\pa p(x)}$
and $f_n^0={\pa  y_3(1,\iota_n)\/\pa p(x)}$,
or $f_n={\pa  y_3(1,\wt\m_n)\/\pa q(x)}$
and $f_n^0={\pa  y_3(1,\iota_n)\/\pa q(x)}$, $x\in[0,1]$.
Then the estimates \er{estvp3-vpa30} and \er{estdoty30y3} yield
\[
\lb{asdotvpdevvp1}
\Big|{f_n\/ y_3(1,\wt\m_n)}
-{f_n^0\/ y_{3,n}^0(1)}\Big|
\le {|f_n-f_n^0|
\/| y_3(1,\wt\m_n)| }
+{|f_n^0| |y_{3,n}^0(1)- y_3(1,\wt\m_n)|\/
| y_3(1,\wt\m_n)| |y_{3,n}^0(1)|}
\le Ce^{-2\nu n}\big(n^2|f_n-f_n^0|+|f_n^0| \|\gu\|_1\big),
\]
for some $C>0$. The identity \er{dotvp30mn} gives
\[
\lb{estdoty30y31}
|\dot y_{3,n}^0(1)|\le {Ce^{2\nu n}\/n^4},
\]
for some $C>0$.
Substituting the estimates \er{asdotvp3} and \er{estdoty30y31}
into  \er{asdotvpdevvp1}, we obtain the first estimate in \er{asdotvpdevvp}.
The asymptotics \er{asderltau3} give \er{dottau3-0}.
The identity \er{grvp3p=0} gives
\[
\lb{estdoty30y32}
\Big|{\pa  y_3(1,\iota_n)\/\pa p(t)}\Big|_{\gu=0}\Big|\le {Ce^{2\nu n}\/n^3},\qq
\Big|{\pa  y_3(1,\iota_n)\/\pa q(t)}\Big|_{\gu=0}\Big|\le {Ce^{2\nu n}\/n^4}.
\]
Substituting the estimates \er{esfdervpvp0} and \er{estdoty30y32}
into  \er{asdotvpdevvp1}, we obtain \er{asdervpdevvp}.

The identity \er{idvp2unp}
and the estimate \er{estvp3-vpa30} imply
\[
\lb{estdoty30wty3}
|\wt y_{3,n}^0(1)|\ge {c_1e^{\nu n}\/n^2},\qq
|\wt y_3(1,\wt\m_n)|\ge |\wt y_{3,n}^0(1)|
-|\wt y_3(1,\wt\m_n)-\wt y_{3,n}^0(1)|
\ge {c_2e^{\nu n}\/n^2},
\]
for some $c_1,c_2>0$.
Let $\wt f_n=\dot{\wt y}_3(1,\wt\m_n)$ and $\wt f_n^0=\dot {\wt y_{3,n}^0}(1)$,
or $\wt f_n={\pa\wt  y_3(1,\wt\m_n)\/\pa p(x)}$
and $\wt f_n^0={\pa\wt  y_3(1,\iota_n)\/\pa p(x)}$,
or $\wt f_n={\pa\wt  y_3(1,\wt\m_n)\/\pa q(x)}$
and $\wt f_n^0={\pa\wt  y_3(1,\iota_n)\/\pa q(x)}$, $x\in[0,1]$.
Then the estimates \er{estvp3-vpa30} and \er{estdoty30wty3} yield
\[
\lb{asdotwtvpdevvp1}
\Big|{\wt f_n\/\wt y_3(1,\wt\m_n)}
-{\wt f_n^0\/\wt y_{3,n}^0(1)}\Big|
\le {|\wt f_n-\wt f_n^0|
\/|\wt y_3(1,\wt\m_n)| }
+{|\wt f_n^0| |\wt y_{3,n}^0(1)-\wt y_3(1,\wt\m_n)|\/
|\wt y_3(1,\wt\m_n)| |\wt y_{3,n}^0(1)|}
\le Ce^{-\nu n}\big(n^2|\wt f_n-\wt f_n^0|+|\wt f_n^0| \|\gu\|_1\big),
\]
for some $C>0$. The identity \er{dotvp30mn} gives
$|\dot{\wt y_{3,n}^0}(1)|\le Ce^{\nu n}n^{-4}$.
Substituting this estimate and the estimates \er{asdotvp3}
into  \er{asdotwtvpdevvp1}, we obtain the second estimate in \er{asdotvpdevvp}.

The identity \er{grwtvp3p=0} gives
\[
\lb{estdotwty30y32}
\Big|{\pa \wt y_3(1,\iota_n)\/\pa p(t)}\Big|_{\gu=0}\Big|\le {Ce^{\nu n}\/n^3},\qq
\Big|{\pa \wt y_3(1,\iota_n)\/\pa q(t)}\Big|_{\gu=0}\Big|\le {Ce^{\nu n}\/n^4}.
\]
Substituting the estimates \er{esfderwtvpvp0} and \er{estdotwty30y32}
into  \er{asdotwtvpdevvp1}, we obtain \er{asdervpdevwtvp}.

ii) Substituting the estimates \er{asdotvpdevvp}
and \er{dottau3-0} into the definition
\er{dfegA}, we obtain \er{estL-Lo}.
Using the identity \er{grmult} and substituting
the estimates \er{dertau3-0}, \er{asdervpdevvp}, and \er{asdervpdevwtvp}
into the definition
\er{dfegB}, we obtain \er{estB-Bo}.~\BBox

\subsection{Asymptotics of the norming constants}
We determine asymptotics of the norming constants and theirs
Frech\'et derivatives.

\medskip

\no {\bf Proof of Theorem~\ref{Thnf}.}
The estimate \er{asncr} is proved in \cite[Th~1.3]{BK25}.
The functions $ y_3(1,\wt\m_n(\gu),\gu)$, $\t_3(\wt\m_n(\gu),\gu),n\in\Z_0$,
are analytic on the ball
$\cB_\C(\ve)$.
The identity \er{idncr} yields that the functions
$h_{sn}$ are also analytic.

Let $\gu\in\cB(\ve)$ and let $n\in\N$.
Substituting the asymptotics \er{aspamun}, \er{estL-Lo},  and
\er{estB-Bo}
into the identity \er{idgradhsn},
we obtain
$$
\sup_{t\in[0,1]}\Big|{\pa h_{sn}\/\pa \gu(t)}
-{\pa h_{sn}\/\pa \gu(t)}\big|_{\gu=0}\Big|
\le{C\|\gu\|_1\/n}.
$$
The identity \er{grhsnp=0}
yields \er{asgrhcn} for $n\in\N$.

We prove the estimate for $n<0$.
Let $n\in\N$. The symmetry \er{defnf} gives
$$
h_{s,-n}(\gu+\d\gu)=-h_{s,n}(\gu_*^-+\d\gu_*^-)
=-h_{s,n}(\gu_*^-)-d_{\gu_*^-}h_{s,n}(\d\gu_*^-)+o(\d\gu)
=h_{s,-n}(\gu)-d_{\gu_*^-}h_{s,n}(\d\gu_*^-)+o(\d\gu).
$$
Moreover,
$$
h_{s,-n}(\gu+\d\gu)=h_{s,-n}(\gu)+d_\gu h_{s,-n}(\d\gu)+o(\d\gu).
$$
Comparing these asymptotics, we get
\[
\lb{dhs-ndh*sn}
d_\gu h_{s,-n}(\d\gu)=-d_{\gu_*^-}h_{s,n}(\d\gu_*^-).
\]
The estimate \er{asgrhcn} implies
$$
d_\gu h_{s,n}(\d\gu)=
-2\pi n\Big({\d p_{cn}\/\sqrt 3}+\d p_{sn}\Big)
+\sqrt 3\Big({\d q_{cn}\/\sqrt 3}+\d q_{sn}
\Big)+{\a_n\/n},\qq |\a_n|\le C\|\gu\|_1.
$$
Substituting $\gu_*^-$ instead of $\gu$
we obtain
$$
d_{\gu_*^-}h_{s,n}(\d\gu_*^-)=
-2\pi n\Big({\d p_{cn}\/\sqrt 3}-\d p_{sn}\Big)
-\sqrt 3\Big({\d q_{cn}\/\sqrt 3}-\d q_{sn}
\Big)+{\b_n\/n},\qq |\b_n|\le C\|\gu\|_1.
$$
The identity \er{dhs-ndh*sn} gives
$$
d_\gu h_{s,-n}(\d\gu)=2\pi n\Big({\d p_{cn}\/\sqrt 3}-\d p_{sn}\Big)
+\sqrt 3\Big({\d q_{cn}\/\sqrt 3}-\d q_{sn}
\Big)-{\b_n\/n},
$$
which yields \er{asgrhcn} for $n<0$.~\BBox

\subsection{Asymptotics of the mappings}
We are ready to prove our results about the mappings.

\begin{lemma}
Each of the mappings $h_n,n\in\Z_0$, is real analytic and satisfies
\[
\lb{asgn}
|h_{n}(\gu)-\cF_{n}\gu|\le{C\|\gu\|_1^2\/n},
\]
\[
\lb{asdergn}
|d_\gu h_n(\d\gu)-\cF_n\d\gu|\le{C\|\gu\|_1\/n}\|\d \gu\|,
\]
for all $\gu\in\cB(\ve),\d\gu\in\gH$, $n\in\Z_0$, and for some $C>0$.
\end{lemma}

\no {\bf Proof.} The functions $h_{cn},h_{sn},n\in\Z_0$, are
analytic on the ball $\cB_\C(\ve)$ and real on $\cB(\ve)$,
therefore, $h_n$ are also analytic and real. The definition
\er{defFhi} gives
$$
\cF_{n}\gu
=\ma-1&{1\/\sqrt3}\\1&-\sqrt3\am \ma\wh q_{cn}+{\wh
p'_{sn}\/\sqrt3}\\-\wh q_{sn}+{\wh p'_{cn}\/\sqrt3}\am =\ma-\wh
q_{cn}-{\wh p'_{sn}\/\sqrt3} -{\wh q_{sn}\/\sqrt3}+{\wh p'_{cn}\/3}
\\
\wh q_{cn}+{\wh p'_{sn}\/\sqrt3} +\sqrt3\wh q_{sn}-\wh p'_{cn}\am.
$$
The estimates \er{as3pev} and \er{asncr} give \er{asgn}.
The estimates \er{aspamun} and \er{asgrhcn} yield \er{asdergn}.~\BBox

\medskip

\no {\bf Proof of Theorem~\ref{ThNablagsn}.}
Let $\gu\in\cB(\ve)$.
The estimate \er{asdergn} shows that $d_\gu h-\cF$ is compact, then
 $d_\gu h$ is the Fredholm operator.
The definition \er{defFhi} shows that $\cF$ is invertible, then $d_\gu h$
is a linear isomorphism. Therefore, due to the Inverse Function Theorem, see,
e.g., \cite[App~B]{PT87},
$h$ is a local analytic isomorphism.
Then we may take $\ve>0$ such that $h$ is a bijection between
$\cB_\C(\ve)$ and $h(\cB_\C(\ve))$.
The asymptotics  \er{asgn} give
$$
\|h(\gu)-\cF\gu\|^2=\sum_{n\in\Z_0}|h_n(\gu)-\cF_n\gu|^2
\le\sum_{n\in\Z_0}\Big({C\|\gu\|_1^2\/n}\Big)^2,
$$
for some $C>0$, which yields \er{estg-cF}.~\BBox

\bigskip

\no\small {\bf Acknowledgments.}
Authors were supported by the RSF grant number
23-21-00023, internet page https://rscf.ru/project/23-21-00023/.

\end{document}